\documentclass[journal]{IEEEtran}
\usepackage[bookmarks=false]{hyperref}
\usepackage{amsmath}
\usepackage{amsfonts}
\usepackage{enumitem,kantlipsum}
\usepackage{adjustbox}
\usepackage{graphicx}
\usepackage[table]{xcolor}
\usepackage{booktabs}
\usepackage{threeparttable}
\usepackage{tabulary}
\usepackage{multirow}
\usepackage{units}
\usepackage{tabu}
\usepackage[binary-units=true]{siunitx}
\usepackage{comment}

\def\BibTeX{{\rm B\kern-.05em{\sc i\kern-.025em b}\kern-.08em
    T\kern-.1667em\lower.7ex\hbox{E}\kern-.125emX}}
 
\newcommand{\rev}[1]{{\color{black}#1}}

\begin{document}
\bstctlcite{use_etal}
%
\title{Q-PPG: Energy-Efficient PPG-based Heart Rate Monitoring on Wearable Devices}

\author{Alessio~Burrello, Daniele~Jahier~Pagliari, Matteo~Risso, Simone~Benatti, Enrico~Macii, Luca~Benini,  Massimo~Poncino%

\thanks{A. Burrello, S. Benatti and L.Benini are with the Department of Electrical, Electronic and Information Engineering, University of Bologna, 40136 Bologna, Italy (e-mail: name.surname@unibo.it).}

\thanks{ S. Benatti is also with the Department of Sciences and Methods for Engineering, University of Modena and Reggio Emilia, Italy (e-mail: name.surname@unimore.it).}

\thanks{L. Benini is also with the Department of Information Technology and Electrical Engineering at the ETH Zurich, 8092 Zurich, Switzerland (e-mail: lbenini@iis.ee.ethz.ch).}

\thanks{D. Jahier Pagliari, M. Risso, and M. Poncino are with the Department of Control and Computer Engineering, Politecnico di Torino, Turin, Italy (e-mail: name.surname@polito.it).}

\thanks{E. Macii is with the Interuniversity Department of Regional and Urban Studies and Planning, Politecnico di Torino, Turin, Italy (e-mail: enrico.macii@polito.it).}
\thanks{This   work   was   supported   in   part   by  the EU Grant Bonsapp  (g.a. no. 101015848). We acknowledge
the CINECA award under the ISCRA initiative, for the availability of
high performance computing resources and support.\\
This article has been accepted for publication in IEEE Transactions on Biomedical Circuits and Systems. This is the author's version which has not been fully edited and
content may change prior to final publication. Citation information: DOI 10.1109/TBCAS.2021.3122017\\
© 2021 IEEE. Personal use is permitted, but republication/redistribution requires IEEE permission.
See https://www.ieee.org/publications/rights/index.html for more information.}
}

\markboth{IEEE Transactions on Biomedical Circuits and Systems,~Vol.~XX, No.~X, January~XXXX}%
{A. Burrello \MakeLowercase{\textit{et al.}}: Q-PPG: Energy-Efficient PPG based Heart Rate Monitoring on Wearable Devices}

\maketitle

\begin{abstract}
Hearth Rate (HR) monitoring is increasingly performed in wrist-worn devices using low-cost photoplethysmography (PPG) sensors. 
However, Motion Artifacts (MAs) caused by movements of the subject's arm affect the performance of PPG-based HR tracking.
This is typically addressed coupling the PPG signal with acceleration measurements from an inertial sensor. Unfortunately, most standard approaches of this kind rely on hand-tuned parameters, which impair their generalization capabilities and their applicability to real data in the field. In contrast, methods based on deep learning, despite their better generalization, are considered to be too complex to deploy on wearable devices.

In this work, we tackle these limitations, proposing a design space exploration methodology to automatically generate a rich family of deep Temporal Convolutional Networks (TCNs) for HR monitoring, all derived from a single ``seed'' model. Our flow involves a cascade of two Neural Architecture Search (NAS) tools and a hardware-friendly quantizer, whose combination yields both highly accurate and extremely lightweight models.
When tested on the PPG-Dalia dataset, our most accurate model sets a new state-of-the-art in Mean Absolute Error. Furthermore, we deploy our TCNs on an embedded platform featuring a STM32WB55 microcontroller, demonstrating their suitability for real-time execution. Our most accurate quantized network achieves 4.41 Beats Per Minute (BPM) of Mean Absolute Error (MAE), with an energy consumption of \unit[47.65]{mJ} and a memory footprint of \unit[412]{kB}. At the same time, the smallest network that obtains a MAE $<$ 8 BPM, among those generated by our flow, has a memory footprint of \unit[1.9]{kB} and consumes just \unit[1.79]{mJ} per inference.
\end{abstract}

\IEEEpeerreviewmaketitle

\section{Introduction}
Modern wrist-worn devices include increasingly heterogeneous sensor sets, monitoring movements and vital parameters such as electrodermal activity and heart rate (HR). HR monitoring, in particular, is important for clinical purposes and precise activity tracking. Early wrist-worn HR tracking devices were connected to a separate chest band, equipped with a simple 1-3 leads Electrocardiogram (ECG) sensor. While accurate, this solution was expensive and created discomfort for users in daily life usage. Therefore, in recent years, ECG chest bands have been progressively replaced by cost-effective and more comfortable photoplethysmography (PPG) sensors, which enable the measurement of HR and blood oxygenation (SpO2) directly from wrist-worn devices~\cite{sviridova2015human}. Examples of commercial devices that include this type of sensor are the Apple Watch~\cite{applewatch} and some Fitbit models~\cite{fitbit}.

PPG sensors consist of one or more Light-Emitting Diodes (LEDs) that periodically emit light onto the skin and a photodetector (i.e., a photodiode) that measures the variations of light intensity caused by blood flow~\cite{tamura2014wearable,castaneda2018review}. More specifically, the larger the blood volume variation, the greater the attenuation of the light emitted by the LED, resulting in a lower current output on the photodiode. Therefore, in an \textit{ideal} PPG signal, peaks can be associated with the HR~\cite{troika2014}.

A major source of inaccuracy in PPG sensors is constituted by motion artifacts (MA), i.e., signal artifacts caused by movements of the user arm and hand, which in turn produce variability in the sensor pressure on the skin or ambient light leaking into the gap between the photodiode and the wrist.
Several studies compared ECG chest straps with PPG-based HR tracking systems~\cite{ge2016evaluating,jan2019evaluation}, showing that the former typically obtain better accuracy, especially in the presence of MAs. As a result, the ECG-based solutions are still considered the reference benchmark for wearable HR tracking~\cite{neurosky}.

To overcome this gap, researchers have recently focused on sensor-fusion approaches that integrate PPG with inertial data from accelerometers in order to detect and mitigate the effect of MAs~\cite{troika2014, reiss2019deep}.
Most of these approaches are based on classical signal processing algorithms such as Independent Component Analysis (ICA), Wiener Filters, and Spectral Peak Detection \cite{temko2017accurate, schack2017computationally, chung2018finite,spama2016}. TROIKA~\cite{troika2014} and its evolution, JOSS~\cite{joss2015} are the seminal works in this field. They estimate the noise caused by MAs via adaptive filtering and then apply spectral peak tracking on the PPG signal to detect the heartbeat frequency. A significant shortcoming of these algorithms is that they rely heavily on hand-tuned parameters, leading to a lack of generalization when evaluated on different datasets.

Deep Learning (DL) is relatively less explored for this task,  mainly for two reasons. First, the deployment of DL models on the resource-constrained computational platforms available on wrist-worn devices, typically based on Microcontrollers (MCUs) is far from trivial~\cite{al2018review}. The primary limiting resource is memory, since highly accurate DL models typically involve millions of parameters, which exceed the memory available in most MCUs. Second, DL solutions necessitate large amounts of training data, which were not available for this task until recently. Things improved with the introduction of PPG-Dalia~\cite{reiss2019deep}, a large dataset for PPG-based HR tracking in the presence of MAs, which includes recordings from 15 subjects performing different daily activities.
Indeed, the authors of~\cite{reiss2019deep} also introduced the first DL solution for this task, based on Convolutional Neural Networks (CNNs), which was shown to outperform state-of-the-art algorithms on the new and more challenging dataset. Following their example, other researchers have then proposed different DL models for this task, such as CorNET~\cite{cornet2019} and Binary CorNET~\cite{rocha2020binary}, which combine convolutional and recurrent layers. All these approaches are based on hand-designed \rev{Neural Network (NN)} architectures, which may be sub-optimal in terms of accuracy versus complexity trade-off. Moreover, they also have limited flexibility from the point of view of deployment, since a fixed architecture cannot be easily adapted to hardware targets with different memory, latency or energy constraints.

In this paper, which extends~\cite{risso2021robust}, we propose the first systematic flow to optimize DL models for PPG-based HR tracking. We focus in particular on Temporal Convolutional Networks (TCNs), a family of DL models that are both HW-friendly and accurate for time-series processing.
Our proposed methodology generates a rich set of Pareto-optimal TCNs in the accuracy versus memory (or n. of operations) design plane, among which designers can then select a particular model based on the constraints of their hardware target.
In detail, the following are the main novel contributions of this work:
\begin{itemize}
    \item We leverage Neural Architecture Search (NAS) to obtain Pareto-optimal TCN architectures that predict a user's HR based on raw PPG and acceleration data. All TCNs are automatically derived from a single seed architecture~\cite{zanghieri2019robust}. With respect to~\cite{risso2021robust}, which only optimized the number of feature maps in each TCN layer, in this work, we extend the search also to consider the \textit{dilation} parameter of convolutional layers, which effectively reduces the model complexity with a limited impact on accuracy.
    \item After optimizing the model architectures, we perform a further search step to select the best data representation format for the networks' parameters and intermediate input/outputs. This hardware-friendly \textit{quantization} enables further model size reductions, thus enriching and improving the Pareto frontier.
    \item We deploy the models resulting from our search on a real embedded platform with a smartwatch form-factor~\cite{polonelli2021hwatch}. The platform includes a STM32WB55 MCU from ST Microelectronics, based on an ARM Cortex-M4 MCU, and the MAX30101 PPG sensor. Furthermore, we also discuss how the results of our flow would change for other, more memory-constrained targets.
\end{itemize}

On PPGDalia, the best performing model obtained with our flow,
%
%
coupled with a simple smoothing post-processing, achieves a Mean Absolute Error (MAE) of 4.36 BPM, and includes $\approx$ 269k trainable parameters. With an additional fine-tuning step, the MAE is further reduced to \unit[3.61]{BPM}.
%
%
After quantization and deployment on the STM32WB55, the smallest model with a MAE $<$ 8 BPM and the most accurate one consume \unit[1.79]{mJ} and \unit[47.65]{mJ} per inference, with a latency of \unit[71.6]{ms}, and \unit[1.9]{s}, and an error of \unit[7.73]{BPM} and \unit[4.41]{BPM}, respectively. 
These two models are respectively 32154.3-145.63$\times$ smaller and require 3711.1-19.6$\times$ fewer operations per inference compared to the previous state-of-the-art DL solution~\cite{reiss2019deep}, while also significantly improving the HR tracking accuracy.

The rest of the paper is organized as follows.
Section~\ref{sec:related} provides an overview of the existing PPG-based HR estimation algorithms.
Section~\ref{sec:background} provides the required background.
Section~\ref{sec:method} describes the proposed optimization methodology,  while Section~\ref{sec:results} presents the experimental results and their discussion. 
Lastly, Section~\ref{sec:conclusions} concludes the paper.

\section{Related Work}\label{sec:related}
\begin{table*}
\centering
\caption{State-of-the-art comparison table. Different MAE results correspond to different datasets. Abbreviations: f.t. $=$ fine-tuning. Bold text points to accuracy on the target dataset, PPG-Dalia.}
\label{tab:SoA}
\begin{adjustbox}{max width=\textwidth}
\begin{tabular}{|c|c|c|c|c|c|c|c|}
\hline
Work                                              & Dataset                                                                                             & Activities                                                                            & Sign.                                              & Pre-Processing                                                                                 & Algorithm                                                                                          & Post-Proc.                                                              & MAE                                                           \\ \hline \hline
\multicolumn{8}{|l|}{\textbf{Classical methods}}               \\ \hline\hline
TROIKA, 2014 \cite{troika2014}                & SPC, 12 subj.                                                                                   & Rest, Running                                                                         & \begin{tabular}[c]{@{}c@{}}PPG, \\ Acc.\end{tabular} & \begin{tabular}[c]{@{}c@{}}0.5-4 Hz filtering, \\ Downsampling\end{tabular}                    & \begin{tabular}[c]{@{}c@{}}Signal decomp., \\ reconstruct., \\ peak tracking\end{tabular} & \begin{tabular}[c]{@{}c@{}}th., \\ hist. track.\end{tabular}            & 2.34 BPM                                                      \\ \hline
JOSS, 2015 \cite{joss2015}                  & SPC, 12 subj.                                                                                   & Rest, Running                                                                         & \begin{tabular}[c]{@{}c@{}}PPG, \\ Acc.\end{tabular} & \begin{tabular}[c]{@{}c@{}}0.5-4 Hz filtering, \\ Downsampling\end{tabular}                    & \begin{tabular}[c]{@{}c@{}}MMV, \\ spectral subtraction\end{tabular}                               & \begin{tabular}[c]{@{}c@{}}th., \\ hist. track.\end{tabular}            & 1.28 BPM                                                      \\ \hline
IMAT, 2015 \cite{mashhadi2015heart}         & SPC, 12 subj.                                                                                   & Rest, Running                                                                         & \begin{tabular}[c]{@{}c@{}}PPG, \\ Acc.\end{tabular} & 0.5-4 Hz filtering                                                                             & \begin{tabular}[c]{@{}c@{}}MA cancellation \\ by SVD, IMAT\end{tabular}                            & \begin{tabular}[c]{@{}c@{}}peak sel., \\ th.\end{tabular}               & 1.25 BPM                                                      \\ \hline
SpaMa, 2016 \cite{spama2016}                 & \begin{tabular}[c]{@{}c@{}}SPC, 12 subj.\\ SPC, 23 subj. \\ Chon Lab, 10 subj.\\ Dalia, 15 subj.\end{tabular} & \begin{tabular}[c]{@{}c@{}}Rest, Running, \\ Rehab. ex.,\\ Rest, Running \\ 8 daily activities \end{tabular} & \begin{tabular}[c]{@{}c@{}}PPG, \\ Acc.\end{tabular} & \begin{tabular}[c]{@{}c@{}}0.5-3 Hz filtering, \\ Downsampling\end{tabular}                    & \begin{tabular}[c]{@{}c@{}}spectral filtering \\ based on PSD\end{tabular}                         & \begin{tabular}[c]{@{}c@{}}hist. track., \\ spline interp.\end{tabular} & \begin{tabular}[c]{@{}c@{}}0.89 BPM\\ 3.36 BPM \\1.38 BPM \\ \textbf{11.06 BPM}\end{tabular}   \\ \hline
WFPV, 2017 \cite{temko2017accurate}         & \begin{tabular}[c]{@{}c@{}}SPC, 12 subj.\\ SPC, 23 subj.\end{tabular}                       & \begin{tabular}[c]{@{}c@{}}Rest, Running, \\ Rehab. ex.\end{tabular}                  & \begin{tabular}[c]{@{}c@{}}PPG, \\ Acc.\end{tabular} & \begin{tabular}[c]{@{}c@{}}0.5-4 Hz filtering, \\ z-score scaling,\\ Downsampling\end{tabular} & \begin{tabular}[c]{@{}c@{}}Wiener filtering, \\ phase vocoder\end{tabular}                         & \begin{tabular}[c]{@{}c@{}}th., \\ hist. track.\end{tabular}            & \begin{tabular}[c]{@{}c@{}}1.02 BPM\\ 1.97 BPM\end{tabular}   \\ \hline
Schack2017 \cite{schack2017computationally} & \begin{tabular}[c]{@{}c@{}} SPC, 12 subj.\\ Dalia, 15 subj. \end{tabular}                                                                                  & \begin{tabular}[c]{@{}c@{}}Rest, Running, \\ 8 daily activities \end{tabular}   & \begin{tabular}[c]{@{}c@{}}PPG, \\ Acc.\end{tabular} & \begin{tabular}[c]{@{}c@{}}0.5-6 Hz filtering, \\ Downsampling\end{tabular}                    & \begin{tabular}[c]{@{}c@{}}Corr.-based Freq. \\ indicating func., \\ FFT\end{tabular}              & th.                                                                     & \begin{tabular}[c]{@{}c@{}}1.32 BPM \\\textbf{ 20.5 BPM}\end{tabular}                                                     \\ \hline
FSM, 2018 \cite{chung2018finite}           & SPC, 23 subj.                                                                                    & \begin{tabular}[c]{@{}c@{}}Rest, Running, \\ Rehab. ex.\end{tabular}                  & \begin{tabular}[c]{@{}c@{}}PPG, \\ Acc.\end{tabular} & \begin{tabular}[c]{@{}c@{}}0.5-4 Hz filtering, \\ z-score scaling,\\ Downsampling\end{tabular} & Winer filtering                                                                                    & FSM                                                                     & 0.99 BPM                                                      \\ \hline
CurToSS, 2020 \cite{zhou2020heart}           & \begin{tabular}[c]{@{}c@{}}SPC, 12 subj.\\ SPC, 23 subj. \\ Dalia, 15 subj.\end{tabular} & \begin{tabular}[c]{@{}c@{}}Rest, Running, \\ Rehab. ex.,\\ 8 daily activities\end{tabular}                  & \begin{tabular}[c]{@{}c@{}}PPG, \\ Acc.\end{tabular} & \begin{tabular}[c]{@{}c@{}}0.5-4 Hz filtering\end{tabular} & \begin{tabular}[c]{@{}c@{}}SSR \\ Curve tracking\end{tabular}  & N/A                                                              & \begin{tabular}[c]{@{}c@{}}2.2 BPM \\ 4.5 BPM \\ \textbf{5.0 BPM}\end{tabular}   \\ \hline
TAPIR, 2020 \cite{huang2020robust}           & \begin{tabular}[c]{@{}c@{}}SPC, 12 subj.\\ SPC, 23 subj. \\ Dalia, 15 subj.\end{tabular} & \begin{tabular}[c]{@{}c@{}}Rest, Running, \\ Rehab. ex.,\\ 8 daily activities\end{tabular}                  & \begin{tabular}[c]{@{}c@{}}PPG, \\ Acc.\end{tabular} & \begin{tabular}[c]{@{}c@{}}0.5-4 Hz filtering\end{tabular} & \begin{tabular}[c]{@{}c@{}}Adaptive filter \\ Peak detection \\ Linear Transform.\end{tabular} & Notch filter                                                                     & \begin{tabular}[c]{@{}c@{}}2.5 BPM \\ 5.9 BPM \\ \textbf{4.6 BPM}\end{tabular}    \\ \hline\hline

\multicolumn{8}{|l|}{\textbf{Deep Learning}}      \\ \hline\hline
DeepPPG, 2019 \cite{reiss2019deep}               & \begin{tabular}[c]{@{}c@{}}SPC, 12 subj.\\ Dalia, 15 subj.\end{tabular}                     & \begin{tabular}[c]{@{}c@{}}Rest, Running,\\ 8 daily activities\end{tabular}           & \begin{tabular}[c]{@{}c@{}}PPG, \\ Acc.\end{tabular} & \begin{tabular}[c]{@{}c@{}}STFT,\\ 0-4 Hz filtering\end{tabular}                               & CNN                                                                                                & N/A                                                                     & \begin{tabular}[c]{@{}c@{}}4 BPM \\ \textbf{7.65 BPM}\end{tabular}     \\ \hline
CorNET, 2019 \cite{cornet2019}                & \begin{tabular}[c]{@{}c@{}}SPC, 12 subj.\\ SPC, 23 subj.\end{tabular}                       & \begin{tabular}[c]{@{}c@{}}Rest, Running, \\ Rehab. ex.\end{tabular}                  & \begin{tabular}[c]{@{}c@{}}PPG\end{tabular}  & \begin{tabular}[c]{@{}c@{}}0.4-18 Hz filtering,\\ z-score scaling\end{tabular}                 & CNN+LSTM                                                                                           & N/A                                                                     & \begin{tabular}[c]{@{}c@{}}4.67 BPM  \\ 5.55 BPM\end{tabular} \\ \hline
Binary CorNET, 2020 \cite{rocha2020binary}               & \begin{tabular}[c]{@{}c@{}}SPC, 12 subj.\\ SPC, 23 subj.\end{tabular}                       & \begin{tabular}[c]{@{}c@{}}Rest, Running, \\ Rehab. ex.\end{tabular}                  & \begin{tabular}[c]{@{}c@{}}PPG\end{tabular}  & \begin{tabular}[c]{@{}c@{}}0.4-18 Hz filtering,\\ z-score scaling\end{tabular}                 & Bin. CNN+LSTM                                                                                           & N/A                                                                     & \begin{tabular}[c]{@{}c@{}}6.78 BPM  \\ 7.32 BPM\end{tabular} \\ \hline

NAS-PPG, 2021 \cite{song2021ppg}               & Dalia, 15 subj.         &8 daily activities     & PPG & \begin{tabular}[c]{@{}c@{}}FFT,\\ 0.6-3.6 Hz filtering\end{tabular}                 & CNN+LSTM                                                                                           & N/A                                                                     & \textbf{6.02 BPM} \\ \hline

\textbf{Our Work}                                          & \begin{tabular}[c]{@{}c@{}} Dalia, 15 subj.\end{tabular}                     & \begin{tabular}[c]{@{}c@{}} 8 daily activities\end{tabular}           & \begin{tabular}[c]{@{}c@{}}PPG, \\ Acc.\end{tabular} & \begin{tabular}[c]{@{}c@{}}0.5-4 Hz filtering\end{tabular}                               & \begin{tabular}[c]{@{}c@{}}TCN\end{tabular}  & \begin{tabular}[c]{@{}c@{}}th, \\ finetuning\end{tabular}                                                                  & \begin{tabular}[c]{@{}c@{}}\textbf{4.36 BPM} \\ \textbf{+f.t: 3.61 BPM} \end{tabular}                                        \\ \hline
\end{tabular}
\end{adjustbox}
\end{table*}

The study of HR monitoring solutions based on wearable devices equipped with PPG sensors has attracted significant research efforts from academia and industry in recent years.
While tracking the HR is a relatively easy and already solved problem for \textit{steady} subjects, movements usually impair the task's performance, adding noise to the PPG signal.
The main challenge is therefore trying to keep a sufficiently high accuracy, typically measured as the MAE \rev{between the predicted $HR_{pred}$ and ground truth $HR_{true}$, where} $MAE = |HR_{true}-HR_{pred}|$, also during activities with strong movements. \rev{This has to be done under the tight memory, latency and energy constraints of battery-operated wrist-worn devices.
The latter usually have operating frequencies in the order of 10s of MHz and a power envelope lower than 100 mW, thus requiring very low-complexity algorithms for real-time execution.}

Recent algorithms can be split into two main categories. One group includes classical model-driven approaches, based on either time- or frequency-domain extracted features, followed by a series of filtering and peak detection/position-refinement steps. The other group consists of data-driven algorithms, mostly based on deep learning. Fewer works are included in this group, since differently from other fields, where it has become the de facto standard, DL is still relatively unexplored for PPG-based HR monitoring.
Table~\ref{tab:SoA} provides an overview of the main solutions proposed in the literature. 

Starting from classical approaches, the seminal work of~\cite{troika2014} paved the way to the algorithmic exploration in this field, introducing the first public PPG dataset called SPC Cup 2015 (SPC hereinafter).
The paper also proposed a three-stage pipeline called TROIKA, comprising i) a signal decomposition step, ii) spectrum estimation, and iii) a final spectral peak tracking.
Tested on SPC, TROIKA achieves a MAE of 2.34 Beats Per Minute (BPM).
The same authors improved their own algorithm in~\cite{joss2015}, proposing JOSS, a JOint Sparse Spectrum reconstruction approach, where spectral difference is used to remove motion artifacts from the PPG spectrum, further reducing the MAE to just 1.28 BPM.
Similar to this approach, in~\cite{mashhadi2015heart}, the authors propose to suppress motion artifacts using Singular Value Decomposition (SVD), coupled with an Iterative Method with Adaptive Thresholding (IMAT), and a final peak selection step. This work slightly improves the performance on the SPC dataset, obtaining a MAE of 1.25 BPM.
Other model-driven approaches \cite{temko2017accurate, schack2017computationally, chung2018finite} use FFT and Wiener filtering to remove motion artifacts from noisy PPG signals, further improving the performance on SPC to 0.99 BPM of MAE.
Until now, the best MAE result on the SPC dataset (0.89 BPM) has been achieved by SpaMa~\cite{spama2016}, a complex five-step pipeline that combines spectral filtering and spline interpolation.

In 2019, Reiss et al. \cite{reiss2019deep} released 
%
%
a new dataset for PPG-based HR monitoring
called PPG-Dalia (Dalia hereinafter), with a higher number of subjects (15) and more activities per subject (8 in total), including daily-life tasks such as driving, sitting or walking.
The most recent model-driven algorithms~\cite{zhou2020heart, huang2020robust} are optimized on this new dataset. The first, CurToSS~\cite{zhou2020heart}, improves JOSS by using sparse signal reconstruction for both acceleration and PPG signals, obtaining 5.0 BPM of MAE. The second, TAPIR~\cite{huang2020robust}, relies on linear temporal transformations, strongly reducing the computational complexity while achieving a MAE of 4.6 BPM.
Noteworthy, all these model-based algorithms include many free parameters, which leads to over-fitting the dataset used for hand-tuning them.
%
%
Therefore, using model-driven algorithms could strongly impair generalization, leading to badly performing solutions in real-life situations, represented in table \ref{tab:SoA} by the challenging DALIA dataset.
Moreover, to the best of our knowledge, none of the aforementioned algorithms has been deployed on wearable devices, probably due to the high complexity of some of the algorithms.
%

In recent years, motivated by the increasing success of deep learning in other bio-signal applications (e.g., gesture recognition~\cite{zanghieri2019robust}, seizure detection~\cite{burrello2020hyperdimensional, burrello2020ensemble} and brain-computer interfaces~\cite{zhang2019survey}), some researchers have started exploring \rev{deep NNs}, in particular \rev{Convolutional Neural Networks (CNNs)} and \rev{Recurrent Neural Networks (RNNs)}, for PPG-based HR tracking.
The seminal work in this group is~\cite{reiss2019deep}, which in conjunction with the publication of Dalia, introduced different CNN variants coupled with a short-time Fourier transform, which outperformed the best model-driven methods~\cite{spama2016,schack2017computationally} on the new dataset.
CorNET~\cite{cornet2019} and its variant for highly constrained devices, BinaryCorNET~\cite{rocha2020binary} have been introduced to reduce model complexity, achieving comparable results to model-driven methods on the SPC dataset using a deep architecture with a CNN front-end and a long-short term memory (LSTM) RNN to combine multiple time samples. 
Finally, a Neural Architecture Search (NAS) approach has been recently applied to the HR tracking problem in~\cite{song2021ppg}, finding a CNN+LSTM network that achieves 6.02 BPM of MAE on Dalia, while also reducing the complexity of the algorithm compared to~\cite{reiss2019deep}, but being still too large for MCU deployment (800k floating-point parameters).
Indeed, \rev{deep NN} models typically have large memory footprints and high computational complexity. Therefore, their deployment on memory-constrained MCUs, with low energy consumption and respecting real-time latency constraints is not trivial.
In our work, we tackle precisely these challenges. To the best of our knowledge, we are the first to i) explicitly create a Pareto frontier of \rev{NNs} in the MAE versus memory (or number of operations) space, from which different models can be selected based on the hardware target's constraints and ii) investigate the embedding of these models into an actual edge device. Furthermore, we achieve the leading accuracy on the DALIA dataset, outperforming all previous state-of-the-art models.

\section{Background}
\label{sec:background}

\subsection{Temporal Convolutional Network}
\label{subsec:tcn} 

Temporal Convolutional Networks (TCNs) are a sub-class of 1D-Convolutional Neural Networks (CNNs) specialized for time-series processing. Recently, TCNs have been shown to outperform RNNs on several tasks, obtaining higher accuracy for the same number of parameters~\cite{bai2018empirical}. The peculiarity of TCNs with respect to standard 1D-CNNs is the use of \textit{causal} and \textit{dilated} convolutions~\cite{bai2018empirical,lea2016temporal}. 
\textit{Causality} constrains the convolution output $\mathbf{y}_{t}$ to depend only on inputs $\mathbf{x}_{\tilde{t}}$ with $\tilde{t} \leq t$. In other words, outputs are computed looking only at past or present (but not future) inputs. \textit{Dilation} inserts a fixed gap $d$ between the input time-steps processed by convolutional kernels (i.e., \textit{filters}). Dilated convolution is beneficial as it permits an increase of the receptive field of the filters on the time axis, without increasing the number of trainable parameters. In summary, the function implemented by a TCN convolutional layer is:

\begin{equation}\label{eq:1d_conv}
\mathbf{y}_t^m = \sum_{i=0}^{K-1} \sum_{l=0}^{C_{in}-1} \mathbf{x}_{t\,s-d\,i}^l \cdot \mathbf{W}_i^{l,m}
\end{equation}
which is repeated $\forall m \in [0, C_{out}-1]$ and $\forall t \in [0, T-1]$. In the formula, \rev{$\mathbf{x} \! \in \! \mathbb{R}^{C_{in} \times T}$} and \rev{$\mathbf{y} \! \in \! \mathbb{R}^{C_{out} \times T/s}$} are the input and output activations, respectively composed by $C_{in}$ and $C_{out}$ channels or features; $T$ is the output length on the time axis, \rev{$\mathbf{W} \! \in \! \mathbb{R}^{C_{out} \times C_{in} \times K} $} the multidimensional array of filter weights, $d$ the dilation factor, $s$ the stride, and $K$ the filter size.
Originally, TCNs have been proposed as fully-convolutional architectures that stacked multiple layers each implementing (\ref{eq:1d_conv})~\cite{bai2018empirical}.
However, more recent implementations also include other elements, such as pooling and fully-connected (FC) layers, which are analogous to those commonly found in standard CNNs~\cite{zanghieri2019robust,ren2020cloud}. In our experiments, we consider TCN architectures that include all these types of layers. 

\subsection{Hardware setup}
\label{sec:hardware_setup}
\begin{figure}[t]
  \centering
  \includegraphics[width=\columnwidth]{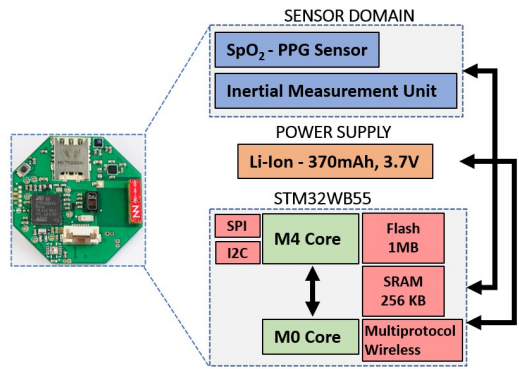}
  \caption{Wrist-worn form factor board presented in \cite{polonelli2021hwatch} used in our experiments and its simplified block diagram.}
  \label{fig:hardware}
\end{figure}
\begin{table}[!t]\caption{Board components power profile.}\label{tab:cpuemodel}
    \centering 
        \begin{tabular}{l|l|l|l}
        Component & State & Current (I) & Power Consumption \\
            \hline\hline
            \multicolumn{4}{l}{\textbf{Microcontroller}}\\\hline
            STM32 & Active & 7.59~mA & 25~mW \\ 
            STM32 & Idle & 4.15~mA & 13.7~mW \\
            STM32 & Stop & $2.45~\mu$A & $8.1~\mu W$ \\
            STM32 & BLE$^*$ & $30~\mu A$ & $99~\mu W$\\
            STM32 & BLE$^{\bowtie}$ & $2.1~mA$ & $6.9~mW$\\
            \hline
            \multicolumn{4}{l}{\textbf{Sensors}}\\\hline
            MAX30101 & Active & $1100~\mu A$ & $5.5~mW$ \\
            MAX30101 & Shutdown & $0.7~\mu A$ & $3.5~\mu W$ \\
            LSM6DS & Active & $9~\mu A$ & $30~\mu W$ \\
            LSM6DS & Shutdown & $3~\mu A$ & $10~\mu W$ \\
            \hline\hline
            \multicolumn{4}{l}{$^*$  STM32 BLE current advertising (0~dBm; 1~s; 31~B).}\\ 
            \multicolumn{4}{l}{$^{\bowtie}$  STM32 BLE  connected master (200~B; 100~ms)} 
        \end{tabular}
\end{table}
We deployed TCNs on the embedded system described in~\cite{polonelli2021hwatch}, whose main board is designed with a wrist-worn form factor. Its picture together with a simplified block diagram of the system is shown in Fig.~\ref{fig:hardware}, where only the components needed for PPG-based HR monitoring are shown.
The board includes a STM32WB55RGV6 System-on-Chip (SoC) from ST Microelectronics~\cite{stm32wb}, referred to simply as STM32WB hereafter. The SoC architecture includes two fully independent cores, an Arm\textsuperscript{\textregistered} Cortex\textsuperscript{\textregistered}‐M4 core running at 64 MHz (application processor) and an Arm\textsuperscript{\textregistered} Cortex\textsuperscript{\textregistered}‐M0+ core at 32 MHz (network processor), optimized for real-time and low-power execution. 
Moreover, the SoC also includes a Radio-Frequency (RF) transceiver with a radio stack compliant with Bluetooth Low Energy 5.0 (BLE) standard, including Bluetooth SIG, Mesh profile, and an HCI for proprietary custom solutions. 
Developed with the same technology of the ultra-low-power STM32L4 MCUs, the STM32WB series provide similar digital and analog peripherals, suitable for applications requiring both extended battery life and high computational capability.

The power supply sub-system of the board exploits a TPS63031 from Texas Instruments, a buck-boost DC/DC converter specifically designed to provide stable output voltage also with impulsive and non-reliable power sources, such as energy harvesters and solar panels. The converter reaches 90\% efficiency during sensor acquisition and processing modes. The local energy buffer exploits a Li-Ion 370~mAh battery used by the TPS63031 as the primary source of power.

The other two relevant components of the system of~\cite{polonelli2021hwatch} for our target applications are two sensors: the MAX30101 \cite{max30101} and the LSM6DSM \cite{lsm6dsm}. The former is a low-power pulse oximeter and PPG module, while the latter is a 6-axes Inertial Measurement Unit (IMU). The two sensors are connected with the MCU using respectively I2C and SPI digital busses.
The hardware power consumption of the different components in all the respective working states, measured through a source/measurement unit Keysight B2900A, is reported in Table~\ref{tab:cpuemodel}. 
The MAX30101 sensor requires a dedicated 5~V for internal LEDs, generated using a step-up converter, with an efficiency of 80\%.

\section{Q-PPG Exploration Flow}\label{sec:method}

\begin{figure*}[ht]
 \centering
\includegraphics[width=.8\textwidth]{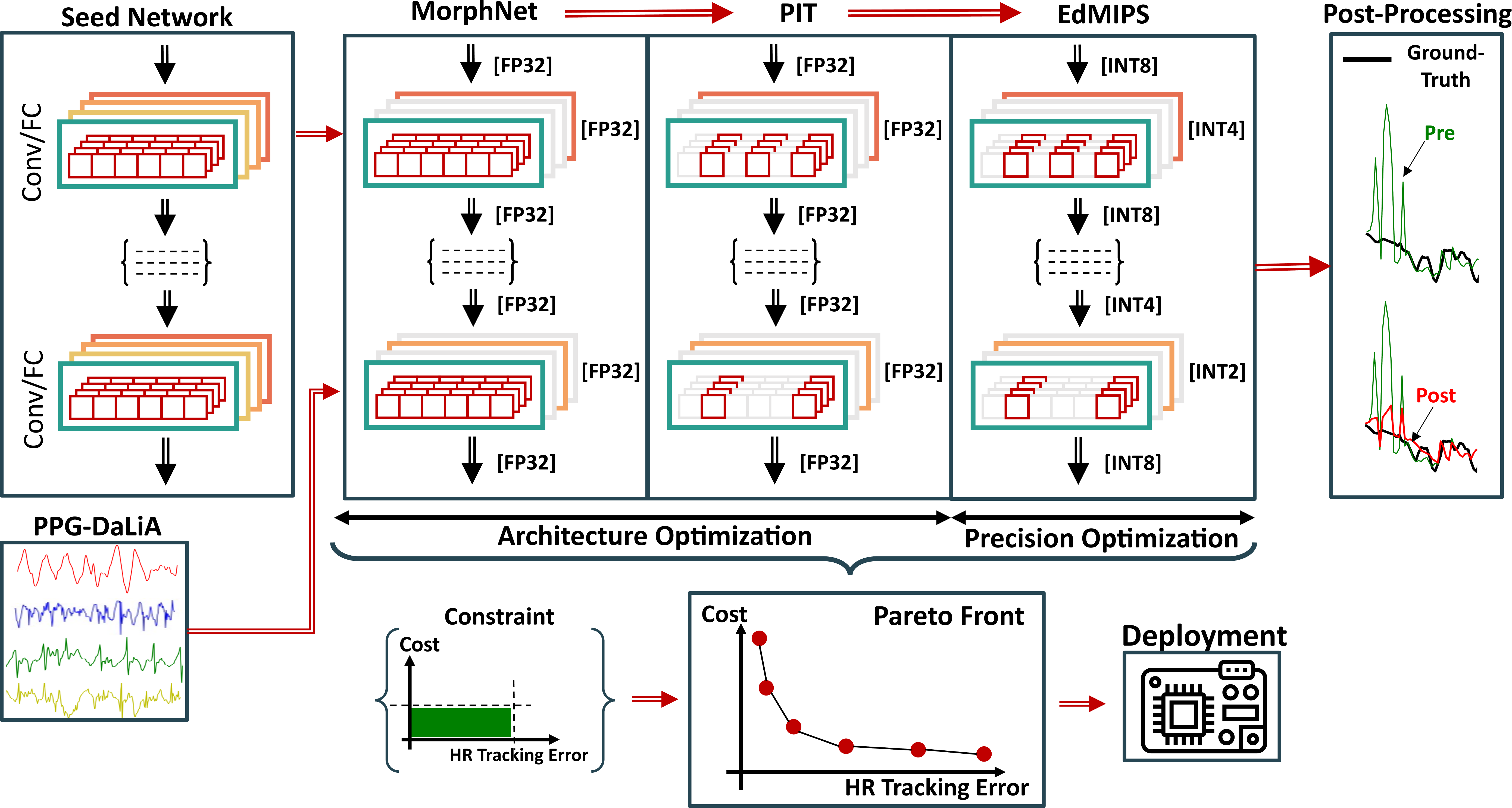}
  \caption{Proposed Q-PPG design space exploration flow.}
  \label{fig:flow}
\end{figure*}

The main contribution of this work is a design space exploration flow able to generate a rich set of HR tracking models, offering diverse trade-offs in terms of  MAE and computational cost, where the latter is measured in terms of number of trainable parameters, or number of operations per inference. We select TCNs as target model type, due to their good performance on time-series processing, and in particular on similar bio-signal processing tasks~\cite{zanghieri2019robust}. 

The inputs of our flow are a training dataset, containing PPG and inertial data associated with the corresponding HR label, and a so-called \textit{seed} TCN, i.e., a sort of ``template'' from which all output models are generated. The flow is then composed of two main phases:
\begin{enumerate}
    \item \textit{Architecture Optimization}: in this phase, we leverage Neural Architecture Search (NAS) tools to explore some of the most important hyper-parameters of the seed TCN to trade-off computational cost and performance.
    \item \textit{Precision Optimization}: in this phase, we further enrich and improve the Pareto curve by applying different types of quantization~\cite{choi2018pact} to the weights and activations of the TCNs produced in phase 1.
\end{enumerate}
At runtime, a low-cost post-processing step is applied to the TCNs produced by phase 2, to further improve their HR tracking accuracy. 
A high-level diagram of the entire flow is shown in Figure~\ref{fig:flow}. Since its final output is a set of quantized TCNs, we name our methodology \textit{Quantized-PPG} (Q-PPG).

Importantly, the lowermost part of the picture shows that the Q-PPG exploration has to be performed \textit{only once} for a given dataset and seed model. After that, deploying to a given hardware target reduces to selecting one of the models from the cost versus error Pareto frontier. Specifically, the target platform imposes constraints on the design space, e.g., limiting the maximum number of parameters based on the available memory space. Then, the most accurate Q-PPG model that meets those constraints is selected and deployed. Therefore, generating an entire \textit{family} of models, rather than a single one makes our methodology efficient and flexible, enabling the deployment of optimized HR tracking solutions not only on the platform described in Section~\ref{sec:hardware_setup}, but also on other similar wearable-class systems.

In the rest of this section, we describe in detail the Q-PPG inputs in Section~\ref{sect:arch_opt_seed}, the two exploration phases in Sections~\ref{sec:arch} (phase 1) and \ref{sec:prec} (phase 2), and the post-processing in Section~\ref{sec:post}.

\subsection{Input Data and Seed Network}\label{sect:arch_opt_seed}

The Q-PPG exploration phase and the training of the final TCNs use the same input dataset, which is composed of raw sensor data gathered from the  PPG-sensor and from a tri-axial accelerometer. Training samples passed to the \rev{NNs} are obtained forming sliding windows of length $T$ on the four signals. Therefore, our TCNs take as input a 2-dimensional array  of size $(T, 4)$. The target output for training is the ground truth HR estimate, expressed as a scalar real number in BPM. HR tracking then reduces to a \textit{regression} problem, where the objective of the TCNs training is to approximate this ground truth value. 
More details on the specific characteristic of the dataset used for our experiments are provided in Section~\ref{sec:dataset}. In all training runs, we use the LogCosh loss function to measure the error between the real and predicted HR. LogCosh has been shown to outperform both RMSE and MAE~\cite{neuneier1998train} as a loss function, favoring the convergence near the minimum, thanks to its smoother behavior around that point.

Besides training data, the other input of our flow is the \textit{seed} network. As better detailed in Section~\ref{sec:arch}, all Q-PPG outputs are obtained starting from the seed, varying its structure (or data precision) to trade-off computational cost and HR tracking error. In particular, the Architecture Optimization phase of Q-PPG tries to \textit{reduce/simplify} the seed, while maintaining the MAE as low as possible. Therefore, in order for our flow to cover the entire design space, the starting point should be a relatively large and accurate TCN.
In this work, the seed network is an adapted version of TEMPONet~\cite{zanghieri2019robust}, a 
TCN which shows impressive results on another bio-signal processing task, i.e., EMG-based gesture recognition. With respect to the original paper, the structure of TEMPONet is slightly modified, i) to make it compatible with the HR tracking task and ii) to widen the space explored by Q-PPG.

A first modification for compatibility with the task consists in changing the first layer to match the input array size. In particular, while 1D convolutional networks can deal with \rev{arbitrary} input lengths on the time axis, the number of input \textit{channels} of the first layer must match the one of the dataset (4 in our case). Similarly, the last FC layer of TEMPONet has also been modified, changing the number of units to 1, as required when performing a scalar regression task. \rev{Lastly, the dilation parameters of all convolutional layers in TEMPONet has been set to $d=1$, while the filter size $K$ has been increased to match the original receptive field.} This has been done because one of the network simplifications performed during the Architecture Optimization consists of increasing the dilation of convolutional layers. Thus, setting $d=1$ everywhere in the seed gives maximum freedom to the search algorithm to explore this parameter.

Apart from these modifications, our seed network is identical to the original TEMPONet of~\cite{zanghieri2019robust}, and consists of a modular \textit{feature extractor}, composed of 3 convolutional blocks, followed by a \textit{classifier} with 3 FC layers.
In turn, each convolutional block contains 3 1D-convolutional layers, where the last layer of each block uses a progressively increasing stride of 1, 2, and 4. Moreover, an average pooling layer is inserted at the end of each block to reduce the output length on the time axis. The number of channels in each block is constant, and it equals 32, 64 and 128 for the 1st, 2nd and 3rd block respectively.
All layers use ReLU activations and batch normalization~\cite{ioffe2015batch}.

\subsection{Architecture Optimization}\label{sec:arch}

This section describes the methodology used to generate different TCN architectures for HR tracking in the accuracy vs. complexity space.
As shown at the top of Figure~\ref{fig:flow}, we leverage a cascade of two different Neural Architecture Search (NAS) tools, called MorphNet~\cite{gordon2018morphnet} and Pruning-In-Time (PIT)~\cite{risso2021pit} for this exploration.

NAS tools automatically generate novel NN architectures for a given task, optimizing hyper-parameters such as the depth and the width of the network, the type of layers included, the connections between layers, etc~\cite{tan2019mnasnet, lin2020mcunet, cai2018proxylessnas}.
Most of these tools target complex computer vision tasks, leading to large and computationally-intensive networks, and requiring an enormous number of training iterations. 
Only recently, researchers have started investigating light-weight NAS approaches called \textit{DmaskingNAS}~\cite{gordon2018morphnet,wan2020fbnetv2}, which search for an optimized architecture in a time comparable to that of a \textit{single} training. Both MorphNet and PIT belong to this category. The efficiency of DmaskingNAS tools comes at the cost of a reduction of the search space, namely in terms of the type of hyper-parameters that are explored. Specifically, all generated models are modified versions of a single \textit{seed} network, as anticipated in Section~\ref{sect:arch_opt_seed}. 

\begin{figure}[ht]
 \centering
\includegraphics[width=\columnwidth]{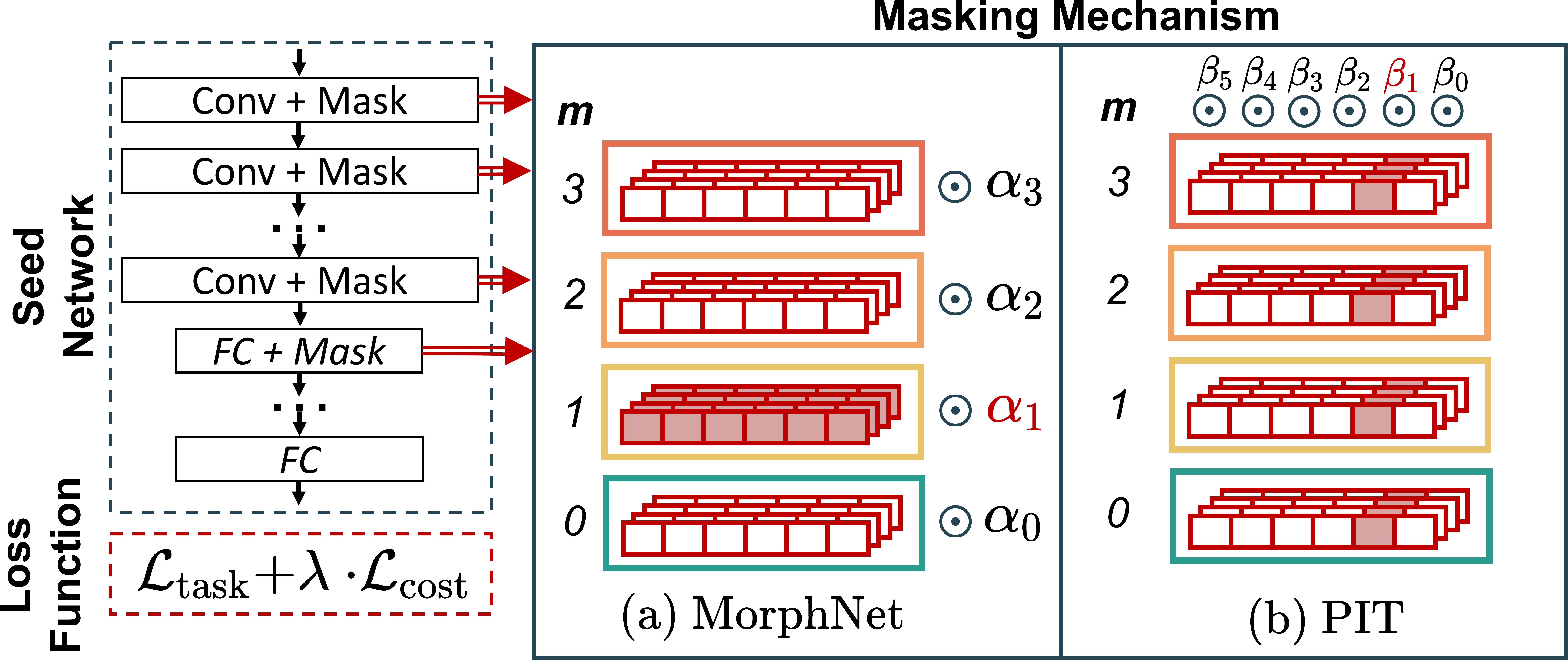}
  \caption{High-level scheme of the functionality of the two NAS algorithms used for architecture optimization. Pooling and other layers are not shown for simplicity.}
  \label{fig:masking}
\end{figure}

A high-level view of the functionality of the two tools used in our work is shown in Figure~\ref{fig:masking}. Before starting the search, the layers of the seed network are modified adding a new set of additional trainable parameters called \textit{masks} ($\alpha_i$ and $\beta_i$ in the figure), each of which multiplies a subset of the layer's weights. In our case, we add masks to convolutional layers of the modified TEMPONet. Moreover, $\alpha_i$ masks (Figure~\ref{fig:masking}a) are also applied to FC layers except the last one. $\beta_i$ masks cannot be applied to FC layers, as explained below.
These masks are then trained together with the normal parameters of the network, encouraging the training algorithm to reduce their magnitude. 

The principle of this approach is that weights multiplied with a small magnitude mask have a negligible impact on the output of a layer (see Eq.~\ref{eq:1d_conv}), and can be removed from the network without increasing significantly its output error. Thus, after training, optimized architectures are obtained by simply eliminating all layer portions corresponding to low magnitude masks.
As shown in Figure~\ref{fig:masking} masks are forced to small values during training by adding to the normal loss function $\mathcal{L}_{\text{task}}$ (i.e., LogCosh for HR tracking) an additional \textit{regularization term} $\mathcal{L}_{\text{cost}}$. The latter computes the expected cost of all the layers of the architecture (e.g., memory occupation or number of operations) as a function of the mask values. Different Pareto points in the complexity versus HR tracking error are obtained changing the relative importance of the two loss terms, through the regularization constant $\lambda$.
%

The two tools used in our work  differ mainly in the masking mechanism. As shown in Figure~\ref{fig:masking}, MorphNet~\cite{gordon2018morphnet} masks all weights relative to the same convolution output channel with one $\alpha_i$. \rev{Therefore, this tool can be used to automatically optimize the \textit{number of output channels} (or features) $C_{out}$ in each convolutional layer.}
In contrast, PIT~\cite{risso2021pit} masks all weights corresponding to the same time-step (and to all output channels) with one $\beta_i$, with the effect of inserting ``holes'' in the convolution filters. \rev{So, this tool can be used to automatically search for the optimal \textit{dilation} parameter $d$ of a TCN.}
To clarify the masking process, all weights multiplied with $\alpha_1$ and $\beta_1$ are coloured in red in Figure~\ref{fig:masking}a and \ref{fig:masking}b respectively. \rev{As shown, $\alpha_1$ is multiplied with all the weights of filter $\mathbf{W}^1$ (i.e., the filter that comprises weights used to derive output channel $C_{1}$), while $\beta_{1}$ is multiplied with the 1-indexed columns of \textit{all} filters, assuming that the latter are stored in channel-major order.}

In the above description, several important details are skipped for sake of space. For instance, MorphNet embeds the masks in the pre-existing parameters of Batch Normalization layers, which are typically placed just after convolutions. Moreover, the training phase aimed at reducing $\alpha_i$s is alternated with \textit{expansion steps} in which the number of channels is uniformly increased in all layers, to recover from the obvious performance penalty. PIT, instead, uses a mechanism to ensure that only \textit{regular} dilation values are obtained from the search, i.e., that the gap between convolution inputs is constant, by imposing relationships among the $\beta_i$s. Further, it binarizes the $\beta_i$s to prevent the training algorithm from compensating small mask values with larger weights. For a detailed discussion on the two tools, readers can refer to the original papers~\cite{gordon2018morphnet,risso2021pit}.

\subsubsection{Search Protocol}

We select MorphNet an PIT for our architecture optimization because both the number of channels and the dilation are key parameters that influence the accuracy and complexity of TCNs~\cite{bai2018empirical}. \rev{However, it has not been previously analyzed how to combine the exploration of these two parameters (N. of channels $C_{out}$ and dilation $d$).} In our experiments, we found empirically that running MorphNet first, followed by PIT, yields much better results than the opposite ordering. Intuitively, this happens because MorphNet operates in a wider and more fine-grained search space, since the possible channels combinations are way more than the possible regular dilation values in a typical convolutional layer.

Given this observation, we use the following search protocol in our work. First, we apply MorphNet to the seed network, with different regularization strengths (from $\lambda = 10^{-6}$ to $\lambda = 10^{-3}$). This results in a first Pareto frontier, composed of TCNs with different number of channels and dilation fixed at 1.
%
%
Then, we select some key points from this frontier, namely the two extremes of the curve (i.e., the TCN achieving the minimum HR tracking error on the validation set and the one with the lowest cost), plus two intermediate solutions. Lastly, we use each of these 4 networks as seeds for PIT, once again repeating the training with different regularization strengths (from $\lambda = 10^{-9}$ to $\lambda = 5 \cdot 10^{-3}$). Consequently, the output of the MorphNet + PIT chain includes 4 (in general, $n$) sets of TCNs, which are then combined to obtain the final Pareto front.
Each NAS execution is preceded by a warm-up phase and followed by a fine-tuning, where only the weights of the seed/optimized TCN are trained. Both these phases have been shown to significantly improve the search quality~\cite{gordon2018morphnet,risso2021pit}. 

\subsection{Precision Optimization}\label{sec:prec}

Starting from the architectures generated by the two cascaded NAS tools, we further expand the space of solutions exploring the per-layer arithmetic precision of our TCNs.
%
%
The quantization technique we use is the same presented in~\cite{capotondi2020cmix}, which was shown to maintain high accuracy even with sub-byte precision, while also being hardware friendly. In fact, differently from other techniques such as weight clustering~\cite{han2015deep}, this method allows to replace all floating point multiply-and-accumulate (MAC) operations required for inference with integer MACs, resulting in a more efficient execution and enabling the deployment of the resulting models on hardware without a Floating Point Unit (FPU).
The method implements a \textit{linear quantizer}, which transforms 
%
%
each floating point tensor $\mathbf{t}$ (of either weights or activations), with values in the range
$[\alpha_\mathbf{t}, \beta_\mathbf{t})$ into a $N$-bit integer tensor $\widehat{\mathbf{t}}$ as:
\begin{equation}
    \widehat{\mathbf{t}} = \mathrm{round} \left( \frac{\mathbf{t} - \alpha_\mathbf{t}}{\varepsilon_\mathbf{t}} \right) \label{eq:2} 
\end{equation}
where $\varepsilon_\mathbf{t} = (\beta_\mathbf{t}-\alpha_\mathbf{t}) / (2^{N}-1)$ is the smallest value that can be represented in the quantized tensor.
The entire inference is then performed using only integer data. Specifically, the accumulation in (\ref{eq:1d_conv}) is performed with \texttt{int32} data, so that no overflows occur, and the final result is then re-quantized as described in~\cite{burrello2020dory}.
Batch normalization layers are also stored and processed with \texttt{int32} format.


Quantization can be applied to a NN either post-training~\cite{capotondi2020cmix} or by means of quantization-aware training (QAT)~\cite{choi2018pact, cai2020rethinking}.  
The first approach works acceptably well for \texttt{int8} data. For instance, in~\cite{risso2021robust} we showed that the MAE degradation for HR monitoring when moving from a single-precision floating point format (\texttt{fp32}) to \texttt{int8} was in the 1.26-1.44 BPM range. However, using QAT leads to the recovery of most of the performance loss for \texttt{int8} precision, and to a limited error increase also for sub-byte precision. This comes at an acceptable cost in terms of training time, since QAT can be applied to an already trained floating point model, reaching convergence in a few epochs. The basic principle of QAT is that of simulating the effect of quantization (so-called \textit{fake quantization}) during the forward pass of each training iteration, while maintaining floating point updates during back-propagation. The details of this technique are out of our scope, and readers can refer to~\cite{choi2018pact,cai2020rethinking}.

\begin{figure}[ht]
 \centering
\includegraphics[width=\columnwidth]{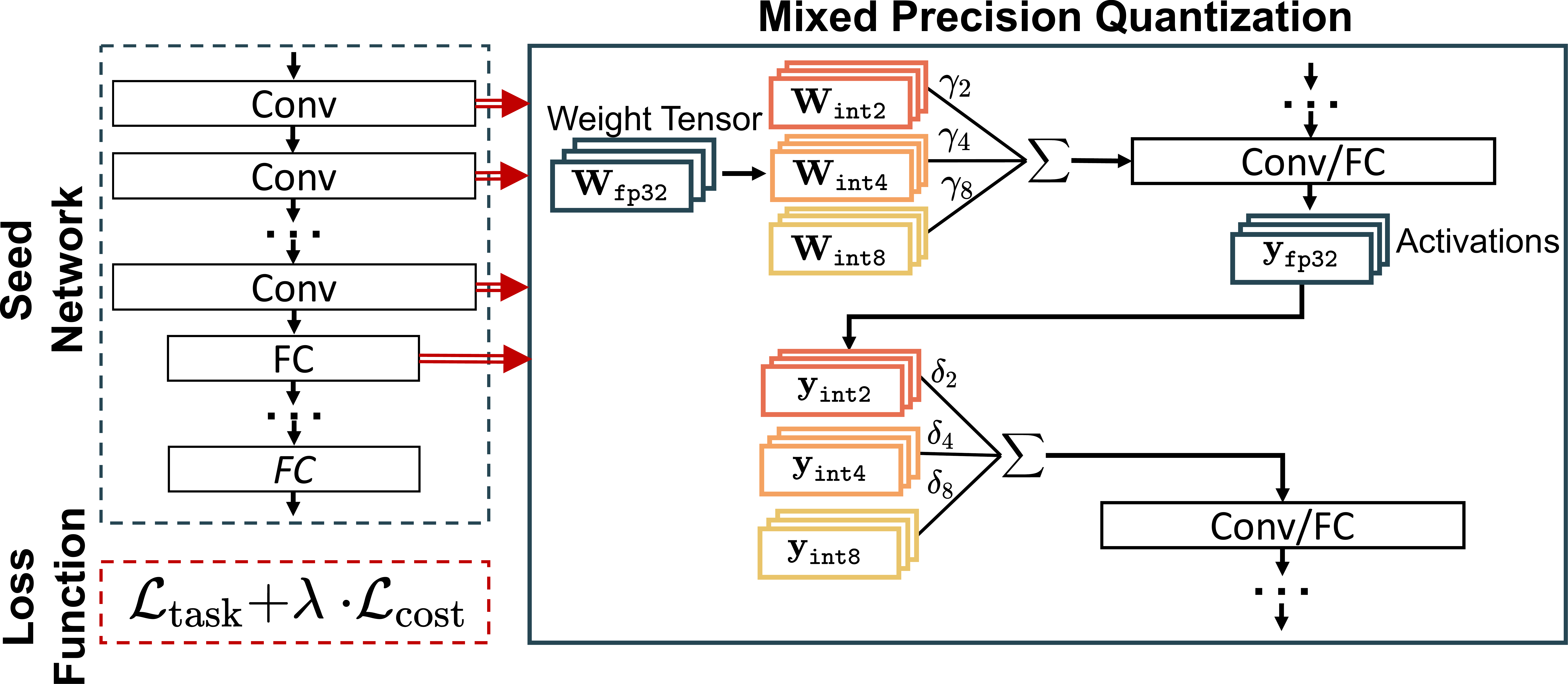}
  \caption{EdMIPS flow for arithmetic precision optimization.}
  \label{fig:edmips}
\end{figure}

In Q-PPG, we use EdMIPS~\cite{cai2020rethinking},
%
%
a tool that allows to simultaneously \textit{i)} perform QAT and \textit{ii)} search for the optimal trade-off among the data format of each layer and the final error of the network.
Figure~\ref{fig:edmips} illustrates the functionality of EdMIPS, which relies on a gradient-based optimization method very similar to the one used by the two NASes described in Section~\ref{sec:arch}.
All convolutional and FC layers in the network are replaced by meta-layers, identical in terms of the executed operation, but whose weights are obtained as combinations of fake-quantized tensors with different precision. For instance,  (\ref{eq:1d_conv}) is changed to:
\begin{equation}\label{eq:1d_conv_q}
\mathbf{y}_t^m = \sum_{i=0}^{K-1} \sum_{l=0}^{C_{in}-1} \mathbf{x}_{t\,s-d\,i}^l \cdot \mathbf{\widehat{W}}_i^{l,m}
\end{equation}
where:
\begin{equation}\label{eq:q_weights}
\mathbf{\widehat{W}} = \sum_{p=0}^{P-1}  \mathbf{Wq}_p \cdot \gamma_p
\end{equation}
and $P$ is the number of different precision formats considered. $\mathbf{Wq}_p$ is the tensor of fake-quantized weights using the $p$-th precision, and  $\gamma_p$ is a trainable coefficient associated to it.
For instance, if we consider \texttt{int2}, \texttt{int4}, and \texttt{int8} formats, then $\mathbf{\widehat{W}} = \mathbf{Wq}_{\texttt{int2}}\cdot \gamma_{\texttt{int2}} + \mathbf{Wq}_{\texttt{int4}}\cdot \gamma_{\texttt{int4}} + \mathbf{Wq}_{\texttt{int8}}\cdot \gamma_{\texttt{int8}}$. All fake-quantized tensors are obtained from the a single, shared, set of floating point weights $\mathbf{W}_{\texttt{fp2}}$.
A similar transformation is also applied to the outputs of the layer, in order to search for the optimal quantization format for activations too. Specifically, the output of the meta-layer is obtained combining fake-quantized activations as follows:
\begin{equation}\label{eq:1d_conv_q_all}
\mathbf{y} = \sum_{p=0}^{P-1} \mathbf{\hat{y}}_p \cdot \delta_p.
\end{equation}
As in the previous NAS approaches, $\gamma$ and $\delta$ coefficients are then trained together with the network weights, adding a secondary loss $\mathcal{L}_{\text{cost}}$ that takes into account the cost of each data format, e.g., the total number of bits required to store the tensor\footnote{The coefficients are passed through a softmax operation, so that they sum to 1, in order not to alter the floating point weights distribution.}.
The training algorithm then assigns a larger coefficient to the fake-quantized tensor that offers the best trade-off between cost and performance. Accordingly, after training, each layer's weights and activations are assigned the data format corresponding to the largest coefficient.

\subsubsection{Search Protocol}

Within Q-PPG, we apply EdMIPS with the following strategy. First, we perform a \textit{uniform} quantization, i.e., using the same bit-width for all tensors ($P=1$), to the entire set of TCNs obtained in the architecture optimization phase. We repeat the QAT with different formats, namely \texttt{int2}, \texttt{int4}, and \texttt{int8}, which are those supported by the backend TCN inference library available for our target hardware~\cite{capotondi2020cmix}. Next, we let the tool search the best bit-width for each tensor, exploring so-called \textit{mixed-precision} networks~\cite{cai2020rethinking}. 
To this end, we select the two extremes of our floating-point Pareto curve, plus two intermediate TCNs with good MAE vs size trade-off, and run EdMIPS with $P=3$, allowing the tool to select among the same three formats listed above. We repeat this search with different regularization strengths $\lambda$ ranging from 10$^{-3}$ to 10$^{-5}$ and merge the results to form the final Pareto front.

\subsection{Post-processing}\label{sec:post}

The last component of our methodology is a post-processing step applied at runtime to the output of our optimized TCNs. This step is orthogonal and independent from the design space exploration described above, and is motivated by the fact that data-driven models such as TCNs, while very accurate on average, may sometimes incur large and unpredictable errors, especially when the processed inputs differ significantly from those seen in the training phase.

Fortunately, in the particular case of HR tracking tasks, some of these errors can be easily filtered out, taking into account the compatibility between TCN estimations and human physiology. Specifically, the dynamics of the human heart rate impose an upper bound on the reasonable variation of the estimate over time, in normal conditions. Therefore, when performing a continuous HR tracking (e.g., every 2s in the experiments described in Section~\ref{sec:results}), a single TCN prediction that differs significantly from all its predecessors is likely due to an error of the model.

\begin{figure}
  \centering
\includegraphics[width=\columnwidth]{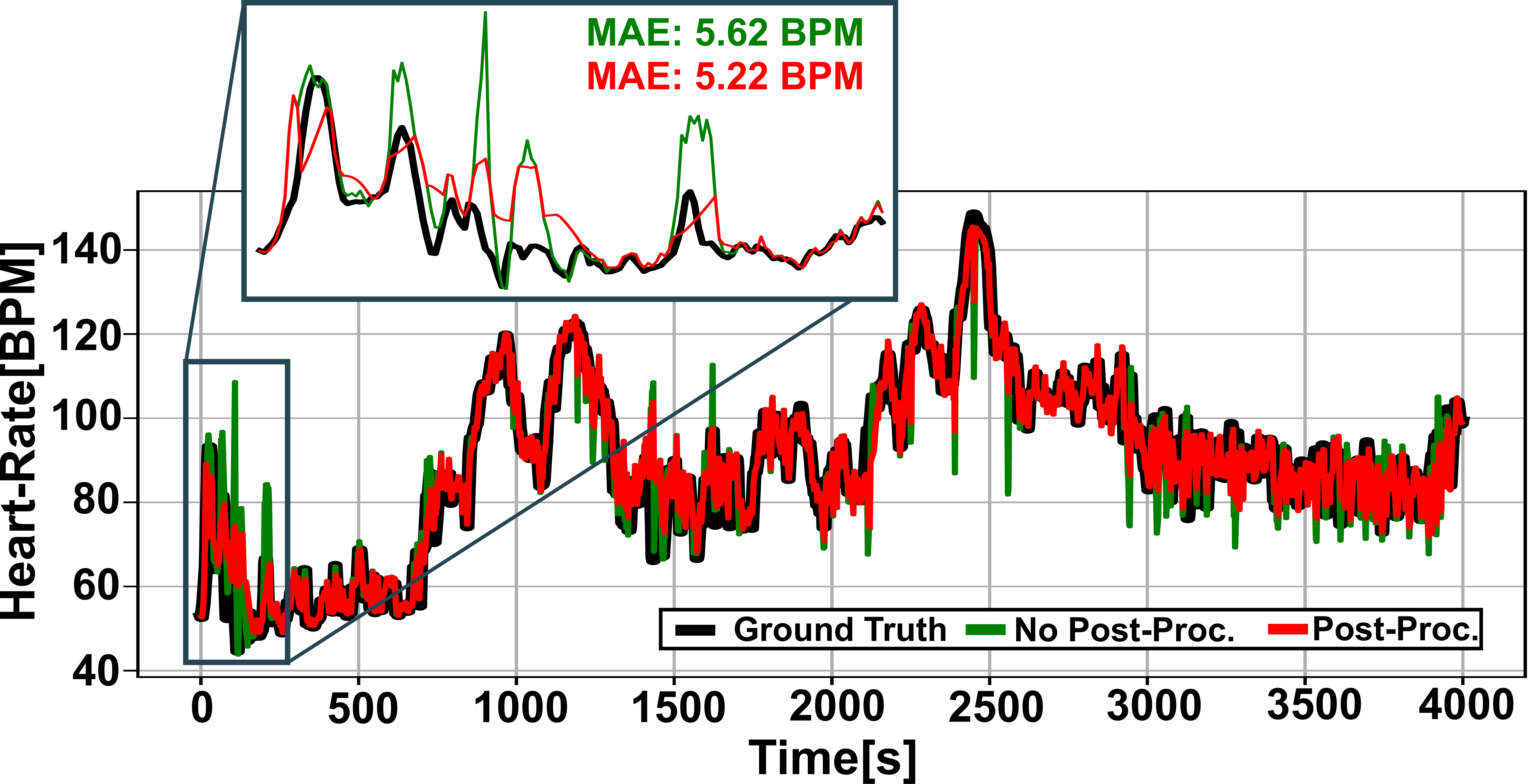}
  \caption{HR tracking obtained with the best performing Q-PPG output TCNs before and after post-processing on the subject n.3 of the Dalia dataset.}
  \label{fig:post_proc}
\end{figure}

Based on these considerations, our post-processing applies a simple filtering on the NN outputs. Specifically, the latest TCN prediction $HR_n$ is compared with the average of the previous N, $E_{n,N} = E[HR_{n-1},...,HR_{n-N}]$. If the difference between these two values is larger than a threshold $P_{th}$, the estimate is \textit{clipped} to $HR_n = E_{n,N}$ $\pm$ $P_{th}$. In this work, we set N to 10 and $P_{th}$ = $\nicefrac{E_{n,N}}{10}$, identical for all patients.
An example of the output produced by one of our models before and after post-processing is shown in Figure~\ref{fig:post_proc}.

\rev{\subsection{Fine-Tuning} 
\label{sec:arch_opt_ft}
In one of our experiments, we also consider partially personalized per-subject models, instead of population ones.
Specifically, after training the models with data of subjects not included in the test-set, we apply a further \textit{fine-tuning} step with a lower learning rate, using the initial 25\% of the data relative to the subject under test. The MAE is then computed on the remaining 75\% of the data.}

\rev{
The main goal of this experiment is to simulate the effect of training a population model on a larger dataset. In fact, despite the fact that PPG-Dalia is the biggest public dataset for PPG-based HR monitoring, it only contains data relative to $<20$ subjects, and the total amount of records is relatively small for training a deep learning model. Therefore, fine-tuning simulates the scenario of a much larger dataset, whose training set contains data that are similar to those seen during testing. Secondarily, this experiment also aims at understanding if \textit{personalized} models could further improve the performance of PPG-based HR monitoring. Collecting ground truth HR labels for personalized models is hard in practice, as it requires dedicated sessions for all subjects, in which both PPG data and ECG references are collected. However, it is not impossible, and it could be beneficial for specific medical use-cases where an extremely accurate model is demanded.}

\section{Results}\label{sec:results}

In this section, we present experimental results to demonstrate the effectiveness of our methodology for building accurate yet efficient HR tracking solutions based on TCNs. Specifically, Section~\ref{sec:dataset} describes our target dataset (PPG-Dalia) and training protocol. Section~\ref{sec:arch_results} shows the results of the \textit{architecture optimization} phase of our flow, which is a set of TCNs with different error and complexity characteristics, but still using floating point data format. These networks are then compared with the state-of-the-art in Section~\ref{sec:soa_comparison}, since all existing algorithms tested on PPG-Dalia use float data. Next, Section~\ref{sec:prec_results} shows how the error and complexity results change after the \textit{precision optimization} phase. Lastly, Section~\ref{subsec:HW_deployment} reports the memory footprint, energy consumption and latency obtained deploying some of the final Q-PPG outputs on the platform described in Section~\ref{sec:hardware_setup}. 
The proposed flow and all TCN training code are implemented in Python 3.6.
To deploy TCNs on the target MCU, we use the open-source Cmix-NN inference library for ARM processors presented in~\cite{capotondi2020cmix}, which supports mixed-precision layers with \texttt{int2}, \texttt{int4} and \texttt{int8} formats for weights and activations.

\subsection{The PPG-Dalia Dataset}\label{sec:dataset}
We evaluate our models as well as state-of-the-art comparisons on the \textit{PPG dataset
for motion compensation and heart rate estimation in Daily-Life Activities} (PPG-Dalia)~\cite{reiss2019deep}. At the time of writing, PPG-Dalia is, to the best of our knowledge, the largest publicly available dataset for PPG-based HR estimation. 
\rev{It includes two PPG channels, from green and red LEDs respectively, (only the former is available in the public version of the dataset), coupled with 3D acceleration data and with the reference ECG signal. In order to compare fairly with previous approaches on this dataset~\cite{reiss2019deep,song2021ppg}, we employ all publicly available data, i.e. the PPG signal from the green LED and the 3D acceleration.}
\rev{The ground truth labels are derived from a chest-worn device with a standard three-points ECG measurement,  using a manually adjusted R-peak detector. Additional activity labels, indicating the type of movement performed by the user, are also available.} The latter are not considered in our experiments. Signals are sampled at 32Hz and organized in sliding windows of 8s ($T$ = 256 samples) with a 6s overlap. The dataset contains a total of 37.5 hours of recording, divided into 15 subjects, eight female and seven male, with age in the interval 21-55.
Two commercial devices were used to collect data: the RespiBAN~\cite{respiban} for the reference ECG, and the wrist-worn Empatica E4~\cite{empatica} for PPG and acceleration data.

\rev{
We validate all models following the cross-validation protocol proposed in~\cite{reiss2019deep}, denoted as Leave-One-Session-Out (LOSO) cross-validation, where the 15 subjects are organized in four randomly picked data folds. Three folds are used as the training set, while the remaining one is subdivided to form the test set, composed of a single subject, and the validation set.
15 training iterations are then performed, each with a different test subject, ensuring that its input data are never used to train the model tested on it.
} 
We compare Q-PPG against both classic and DL methods that have been tested on the same dataset, taking their results directly from the original papers.
When analyzing MCU deployments, we consider a real-time constraint of 2s per inference, equal to the time-shift between two consecutive samples in the dataset, in accordance with previous work~\cite{reiss2019deep, huang2020robust}.

\subsection{Architecture Optimization Results}\label{sec:arch_results}

\begin{figure}[ht]
  \centering
  \includegraphics[width=\columnwidth]{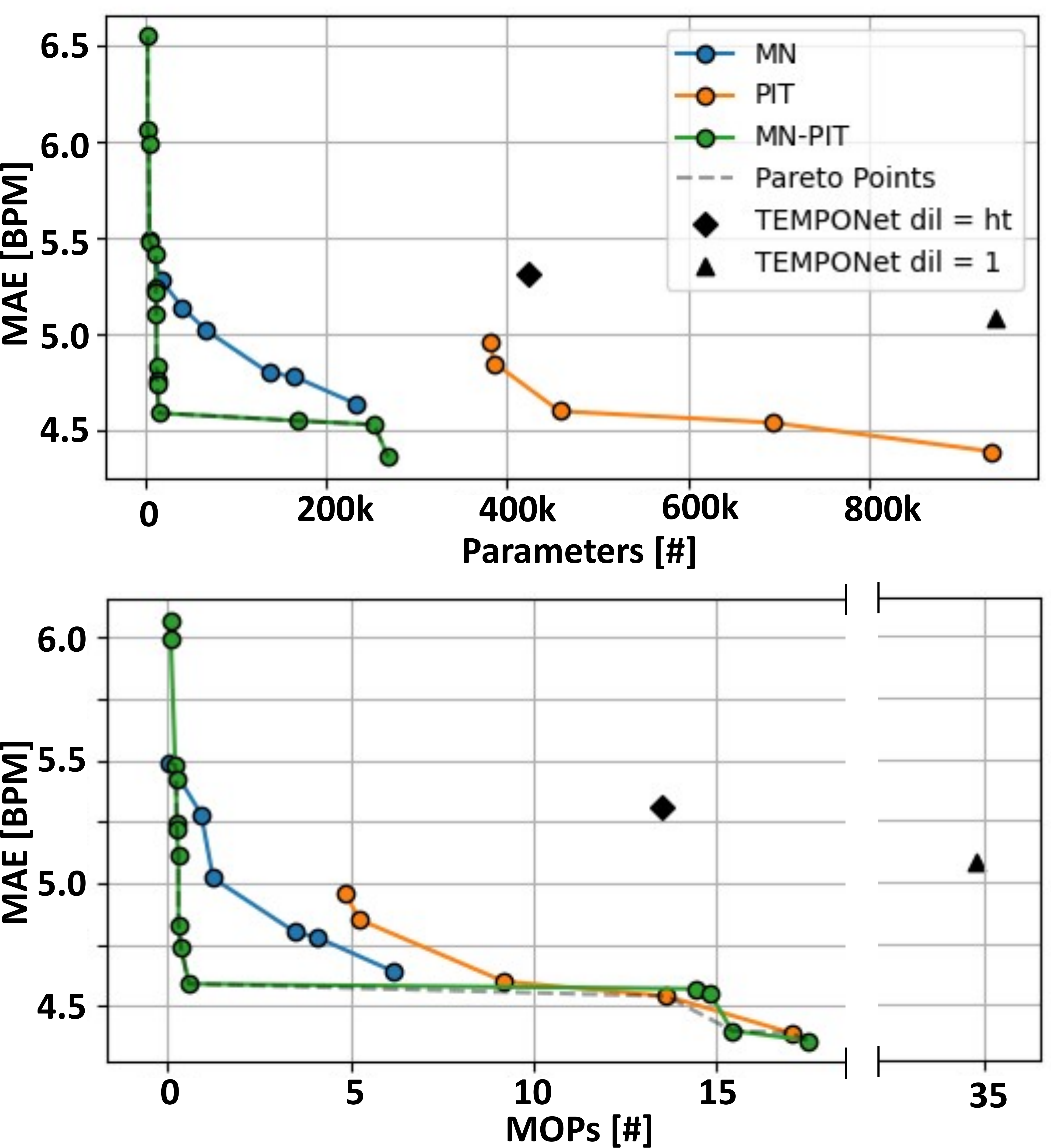}
  \caption{Architecture optimization results in the MAE versus n. of parameters and MAE versus Millions of Operations (MOPs) planes. The curve labeled ``MN-PIT'' corresponds to the sequence of MorhpNet (MN) and PIT used in the proposed Q-PPG flow.}
  \label{fig:arch_results}
  \end{figure}

Figure~\ref{fig:arch_results} reports the results of the architecture optimization phase of our flow. Specifically, it shows (in green) the Pareto frontiers defined by the different TCN variants discovered by Q-PPG, changing the regularization strength of the two NAS algorithms. Results are reported in terms of MAE versus number of trainable parameters and MAE versus number of operations, and include the effect of the runtime post-processing described in Section~\ref{sec:post}. Importantly, these models are still not quantized, and use single-precision floating point representation for both activations and weights. 
  
Besides the outputs of the complete flow, four other results are reported for comparison. The black diamond and triangle correspond respectively to the original TEMPONet, with the dilation of convolutional layers set as in~\cite{zanghieri2019robust}, and to the TEMPONet variant with all dilations set to 1, which corresponds to the input seed of Q-PPG. Comparing these two points with the green curve clearly show that: i) using a hand-tuned TCN originally designed for another task, such as TEMPONet, would be suboptimal and ii) the two NAS algorithms are able to simultaneously improve the HR tracking performance of the seed, while also dramatically reducing its complexity.

\begin{table*}[ht]
\centering
\caption{Comparison with state-of-the-art PPG-based HR monitoring algorithms.}
\label{tab:SoA_comparison}
\begin{tabular}{l|p{0.45cm}p{0.45cm}p{0.45cm}p{0.5cm}p{0.5cm}p{0.5cm}p{0.5cm}p{0.5cm}p{0.5cm}p{0.5cm}p{0.5cm}p{0.5cm}p{0.5cm}p{0.5cm}p{0.5cm}|c}
                        & S1   & S2 & S3   & S4            & S5            & S6            & S7            & S8            & S9            & S10           & S11           & S12           & S13           & S14           & S15           & Mean  \\\hline\hline
\multicolumn{17}{l}{\textbf{Classical Models}}\\\hline\hline                        
Schack2017 \cite{schack2017computationally}          & 33.1         & 27.8         & 18.5         & 28.8         & 12.6         & 8.7          & 20.65         & 21.8         & 22.3         & 12.6         & 21.1         & 22.8         & 27.7         & 12.1         & 16.4         & 20.5           \\ 
SpaMaPlus \cite{spama2016}           & 8.86          & 9.67          & 6.40          & 14.10         & 24.06         & 11.34         & 6.31          & 11.25         & 16.04         & 6.17          & 15.15         & 12.03         & 8.50          & 7.76          & 8.29          & 11.06          \\ 
TAPIR \cite{huang2020robust}    & 4.50          & 4.50          & 3.20          & 6.00          & 5.00          & 3.40          & 2.80          & 6.30          & 8.00          & 2.90 & 5.10          & 4.70         & 3.10          & 5.00          & 4.10          & 4.57        \\ 
CurToSS \cite{zhou2020heart}    & 5.40          & 4.30          & 3.00          & 8.00          & \textbf{2.20} & \textbf{2.80} & 3.30          & 8.50          & 12.60         & 3.60          & 3.60 & 6.10          & 3.00          & 5.50          & 3.70          & 5.04         \\ \hline\hline
\multicolumn{17}{l}{\textbf{Deep Learning Models}}\\\hline\hline         
DeepPPG \cite{reiss2019deep}    & 7.73          & 6.74          & 4.03          & 5.90          & 18.51         & 12.88         & 3.91          & 10.87         & 8.79          & 4.03          & 9.22          & 9.35          & 4.29          & 4.37          & 4.17          & 7.65            \\ 
NAS-PPG \cite{song2021ppg}    & 5.46 & 5.01 & 3.74 & 6.48 & 12.68 & 10.52 & 3.31 & 8.07 & 7.91 & 3.29 & 7.05 & 6.76 & 3.84 & 4.85 & 3.57   & 6.02   \\ \hline

OurWork, Best MAE & 4.29 & 3.62 & 2.44 & 5.73 & 10.33 & 5.26 & \textbf{2.00} & 7.09 & 8.60 & 3.09 & 4.99 & 6.25 & \textbf{1.92} & 3.02 & 3.55 & 4.81 \\ 
+ Post-Processing & 3.78 & 3.36 & \textbf{2.33} & 4.84 & 9.95 & 4.38 & 2.20 & 5.88 & 7.59 & \textbf{2.74} & 4.55 & 5.20 & 2.14 & \textbf{2.99} & 3.47 & 4.36 \\
+ Fine-Tuning & \textbf{3.25} & \textbf{2.55} & 2.66 & \textbf{4.21} & 5.41 & 4.11 & 2.06 & \textbf{5.07} & \textbf{7.15} & 3.04 & \textbf{3.07} & \textbf{3.39} & 2.13 & 3.13 & \textbf{2.96} & \textbf{3.61} \\ \hline
\end{tabular}
\end{table*}

The other curves reported in Figure~\ref{fig:arch_results} show the results of the isolated application of the two NAS algorithms. The blue points refer to the application of MorphNet (MN) to TEMPONet with hand-tuned dilation, and correspond to the results of~\cite{risso2021robust}. Orange points, instead, correspond to the application of PIT alone, using the TEMPONet variant with $d=1$ as seed. Comparing these two curves with the green one clearly shows that combining the two NAS tools is almost always superior with respect to applying them individually, especially when considering MAE versus model size. In fact, the global Pareto frontier (gray dashed line) is almost always overlapped with the Q-PPG output. This result is motivated by the fact that MorphNet alone can explore a wide space of solutions, by tuning the number of channels in each layer, but is forced to use sub-optimal hand-tuned dilation values. In contrast, being unable to alter the number of channels, PIT can only explore a limited portion of the design space. 

Overall, our automatic design space exploration technique is able to span more than two orders of magnitude, both in terms of TCN parameters (3.5k-269k) and  OPs (0.1M-17.5M), despite starting from a single seed TCN.
The most accurate model obtained by our automatic design space exploration, before quantization, achieves a MAE of just 4.36 BPM while requiring around 269k parameters and 17.5M OPs.
On the other hand, by only using MN, as presented in \cite{risso2021robust}, the best MAE obtained was 4.88 BPM, with similar number of parameters (230k) and operations (12M).
\rev{Noteworthy, increasing the number of parameters from 3.5k up to 30k leads to improving the MAE from 6.5 BPM to 4.55 BPM. In comparison, a relatively small improvement of the MAE of 0.19 BPM is obtained by increasing the network dimension of one additional order of magnitude. This could be an indication that, even with relatively few parameters, our NAS-based flow is able to find near-optimal models, that closely approach the best overall performance obtainable with a TCN on this dataset without altering it, e.g., through data augmentation.}

\subsection{State-of-the-art comparison}\label{sec:soa_comparison}
  
\begin{figure}[ht]
\centering
\includegraphics[width=\columnwidth]{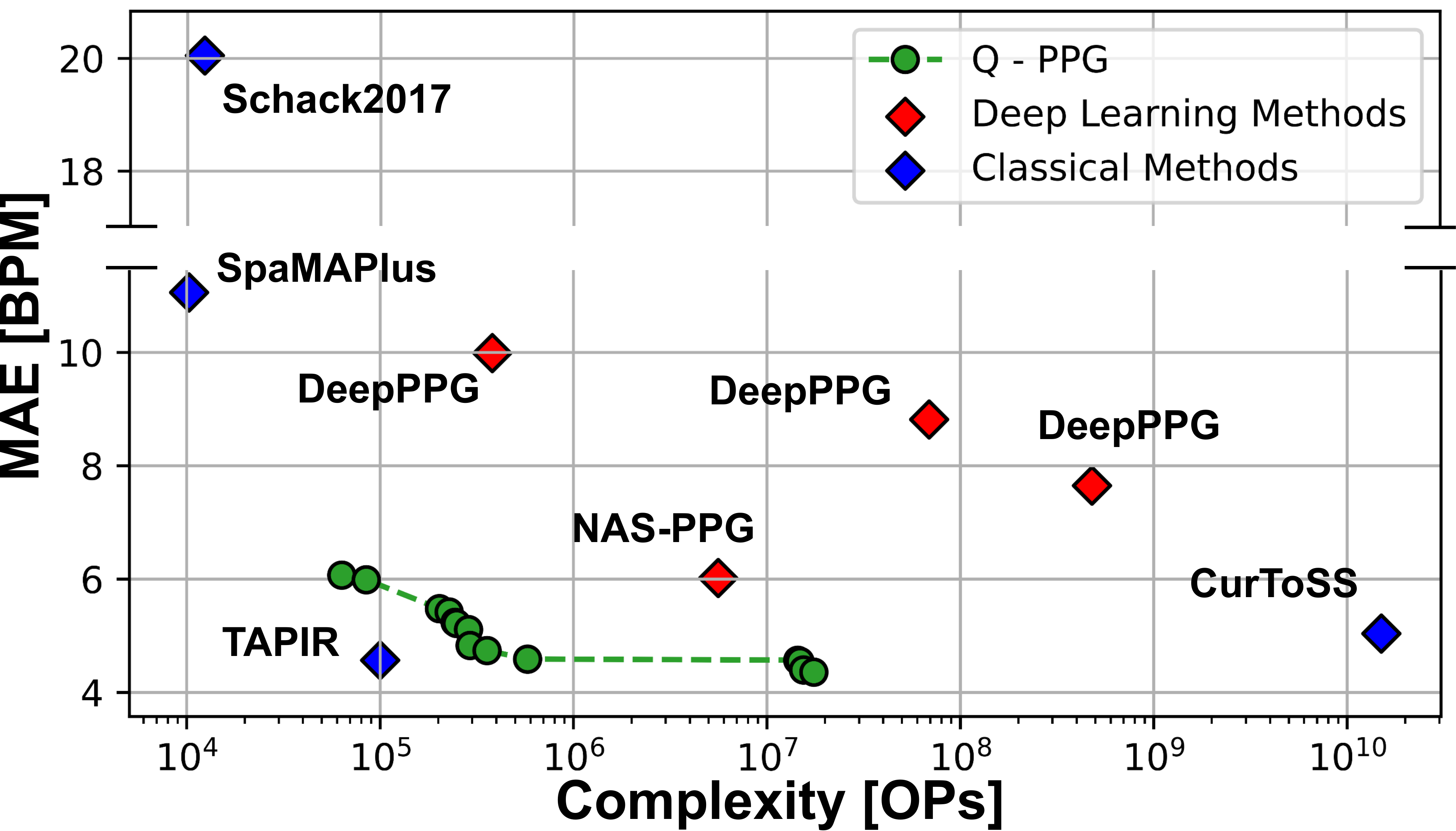}
\caption{Comparison with state-of-the-art algorithms in the MAE versus number of operations space.}
\label{fig:SOA}
\end{figure}

Figure~\ref{fig:SOA} compares our models (in green) with state-of-the-art algorithms, including both classical and deep learning solutions (blue and red points respectively), in the MAE versus number of operations space. For Q-PPG, we report the entire Pareto frontier of TCN architectures, i.e., the same points plotted in the lowermost graph of Figure~\ref{fig:arch_results}. As anticipated, we compare with the state-of-the-art considering non-quantized models, since all previous work use floating point data. The details of the cross-validated MAE results for each of the 15 PPG-Dalia subjects are reported in Table~\ref{tab:SoA_comparison}. For works proposing multiple models (ours and DeepPPG~\cite{reiss2019deep}), the table reports the results of the most accurate one. Q-PPG results are reported both with and without post-processing.

As shown, Q-PPG significantly outperforms \textit{all} previous solutions based on deep learning. With respect to DeepPPG, we achieve a better MAE even with our \textit{simplest} model (6.07 vs 7.65 BPM), despite a striking 7572$\times$ reduction in complexity. With respect to the recent NAS-PPG, the same Q-PPG model obtains comparable MAE (6.07 vs 6.02 BPM) with 88.4$\times$ fewer operations.
Moreover, the best Q-PPG model obtains a MAE that \textit{outperforms the previous state-of-the-art} method for this dataset, TAPIR~\cite{huang2020robust}, although at the cost of higher complexity, achieving an average error of just 4.36 (vs 4.57) BPM.
\rev{TAPIR is not dominated in the Pareto sense due to its very low theoretical complexity. However, it is essential to note that this method has a few shortcomings. First, it has never been deployed on an embedded device, and the optimization of its custom filtering operations for an MCU is far from trivial. Furthermore, the original implementation of TAPIR uses floating-point data, and the effect of integer/fixed-point approximation is not studied. Therefore, even if the target device is equipped with a floating-point unit (which is not always the case for MCUs), the lower number of operations might not actually translate into efficiency benefits with respect to our quantized networks. Second, TAPIR performs poorly with slight parameters modifications, as shown in the original paper. Therefore, as for many other classical methods, the hand-tuning of its parameters is critical and might hamper generality.}

Looking at Table~\ref{tab:SoA_comparison}, it is evident that our best TCN performance is strongly impaired by subject 5. This is due to the fact that this subject's record contain very high HR values, rarely encountered in training data, which are therefore badly predicted by data-driven approaches.
\rev{In fact, if we apply the additional fine-tuning step described in Sec. \ref{sec:arch_opt_ft}, the MAE of the best Q-PPG model further reduces to just 3.61 BPM. 
For instance, for subject 5, we reduce the MAE from 9.95 to 5.41 BPM. Overall, fine-tuning improves the performance of our model on 11 out of 15 subjects.
This demonstrates that our method could achieve potentially even better performance, given the availability of a larger and more varied dataset.}

\subsection{Precision Optimization}\label{sec:prec_results}

\begin{figure}[ht]
  \centering
  \includegraphics[width=\columnwidth]{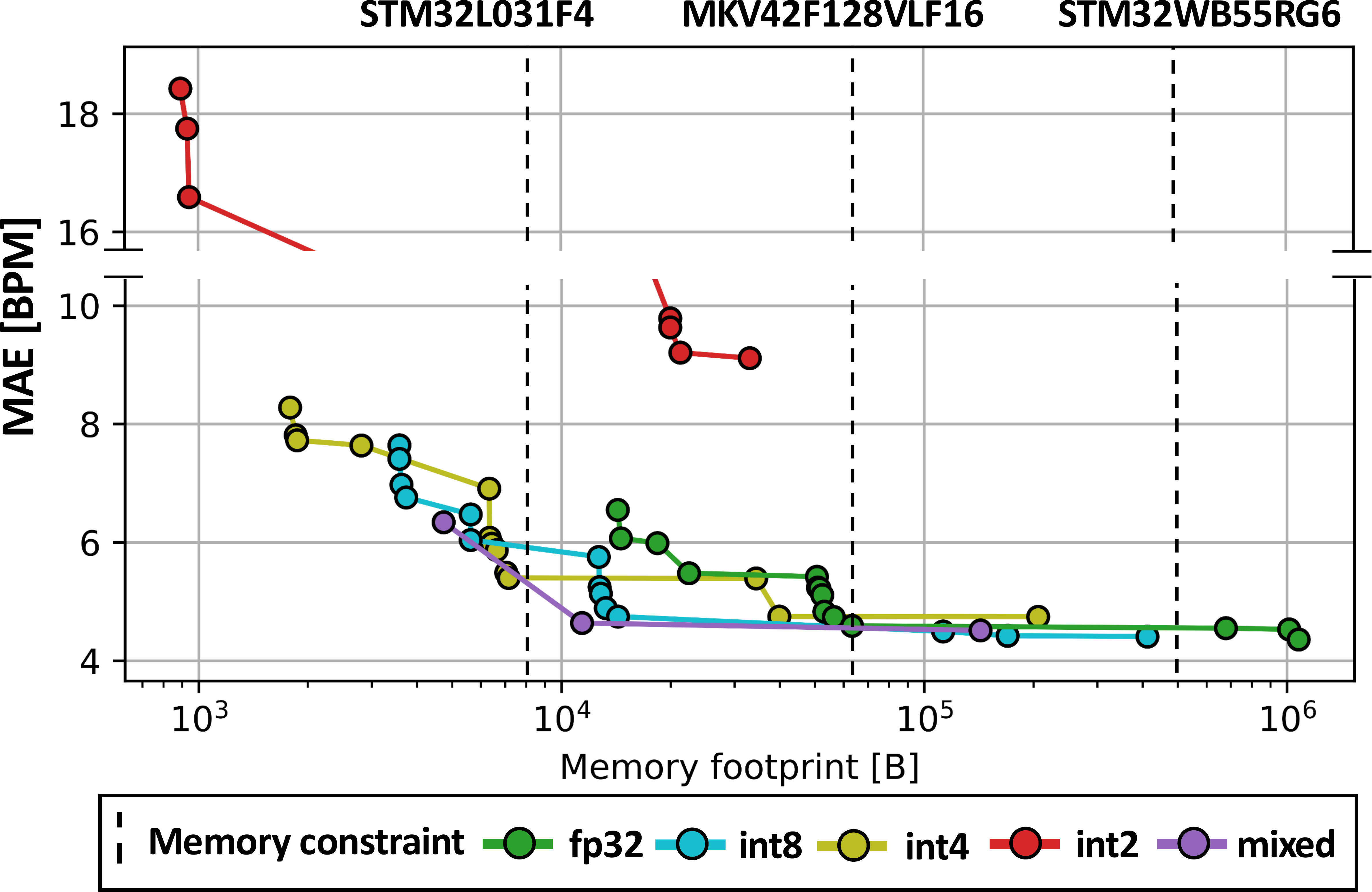}
  \caption{MAE versus memory occupation of Q-PPG TCNs quantized with different data formats.}
  \label{fig:prec_results}
  \end{figure}

Figure~\ref{fig:prec_results} shows how the MAE versus model size results change when applying different types of quantization to Q-PPG models. Note that the x-axis of the curve now reports the model size in bytes, rather than number of parameters, and the dark green curve corresponds to the one in the topmost graph of Figure~\ref{fig:arch_results}. The graph then reports also the results of uniform and mixed-precision quantization, applied with the methods and hardware-friendly formats described in Section~\ref{sec:prec}. 

A first important observation is that uniform int8 quantization incurs very little MAE degradation, as shown by the similar shapes of the dark green and blue curves, despite a reduction of a factor of 4 in model size. Furthermore, both sub-byte uniform quantization (int4 and int2) and mixed-precision results fall on the global Pareto frontier, demonstrating that all formats are useful to obtain different MAE versus model size trade-offs. Overall, thanks to quantization, the range of model sizes obtained with our methodology reaches a span of 3 orders of magnitude, from around 1MB (largest float TCN) to less than 1kB (smallest int2-quantized TCN), with MAE values ranging from 4.36 to $\approx$20 BPM.

To show the flexibility of our flow, the figure also reports, in the form of vertical dashed lines, the constraints imposed by the Flash memory available in 3 different commercial MCUs. The rightmost line corresponds to our target platform (the STM32WB, with 1MB of Flash), whereas the other two correspond, from right to left, to a MKV4 MCU from NXP, based on a Cortex-M4 with 128kB of Flash~\cite{nxpkv4} and to the STM32L031F4, equipped with a Cortex-M0+ and 16kB of Flash~\cite{stm32l0}. Typically, a MCU installed on a wearable device has to store in Flash the code and data for multiple applications. Since the precise application set varies from product to product, here we assumed that 50\% of the total Flash can be devoted to storing the TCN. While the actual constraint may differ in practice, this is just an example for sake of demonstrating a principle that is valid in general. Once the constraint is defined, picking best Q-PPG model for a given hardware simply reduces to finding the most accurate Pareto point left of the corresponding vertical line. In particular, our target STM32WB can fit the most accurate \textit{quantized} model overall, which requires $\approx$412 kB of memory, and achieves a MAE of 4.41 BPM.  In contrast, the largest model fitting the memory of the MKV4 is a mixed-precision TCN requiring just 11.3kB, and achieving an average error of 4.64 BPM. Lastly, for the STM32L0, the most accurate TCN fitting in memory is quantized at 4bit and occupies 7.15 kB, with a MAE of 5.40 BPM.
  
\subsection{Deployment Results}
\label{subsec:HW_deployment}
In this section, we discuss the results obtained deploying three representative TCNs obtained with Q-PPG on the wearable device described in Section~\ref{sec:hardware_setup}.
Specifically, we deployed the smallest networks with less than 8 BPM and 5 BPM of MAE, (Q-PPG-S and Q-PP-M respectively), as well as the most accurate of all quantized networks (Q-PPG-L).
All deployments have been performed using the Cmix-NN layers library~\cite{capotondi2020cmix}, which 
has been adapted to support 1D convolutions with dilation.

The memory occupation, latency and energy consumption of the three networks for a single inference are reported in Table \ref{tab:deployment}, which also shows the type of quantization used by each of them. As expected, smaller models are associated with a larger MAE. However, Q-PPG-M achieves better latency and energy results with respect to Q-PPG-S. This is due to the fact that Q-PPG-M also includes \texttt{int8} layers, which have higher performance than \texttt{int4} in Cmix-NN, since the latter require more complex packing/unpacking operations to match the bit-width of Cortex-M vector ALUs. Interestingly, Q-PPG-M trades-off just 0.23 BPM of MAE for a 36.2$\times$ (34.2$\times$) memory (energy) reduction compared to Q-PPG-L.

%
%

\begin{table}[h]
\centering
\caption{Deployment of different Q-PPG networks on the STM32WB55 using Cmix-NN layers~\cite{capotondi2020cmix}.}
\label{tab:deployment}
\begin{tabular}{l|l|l|l|l}
Model        & Memory [B] & Latency  & Energy   & MAE      \\ \hline \hline
Q-PPG-S (int4)  & 1866        & 71.6 ms  & 1.79 mJ  & 7.73 BPM \\
Q-PPG-M (mixed) & 11388       & 55.7 ms & 1.39 mJ  & 4.64 BPM \\
Q-PPG-L (int8)  & 411997      & 1.90 s   & 47.65 mJ & 4.41 BPM \\ \hline
\end{tabular}
\end{table}
\begin{table}
\centering
\caption{Energy consumption of the three main components of the system during in phases.}
\label{tab:power}
\begin{tabular}{l|p{1.5cm}|p{1.5cm}|p{1.5cm}}
               & MAX30101 {[}mW{]} & LSM6DSM {[}mW{]}  & STM32WB {[}mW{]} \\ \hline \hline
Inference             & 5.5 {[}18.0\%{]}  & 0.03 {[}0.1\%{]} & 25.0 {[}81.9\%{]}    \\
Data Comm.   & 5.5 {[}28.6\%{]}  & 0.03 {[}0.2\%{]}  & 13.7 {[}71.2\%{]}    \\
Data Gath. & 5.5 {[}99.3\%{]}  & 0.03 {[}0.5\%{]} & 0.008 {[}0.2\%{]}    \\ \hline
\end{tabular}
\end{table}
\begin{figure}[ht]
  \centering
  \includegraphics[width=0.9\columnwidth]{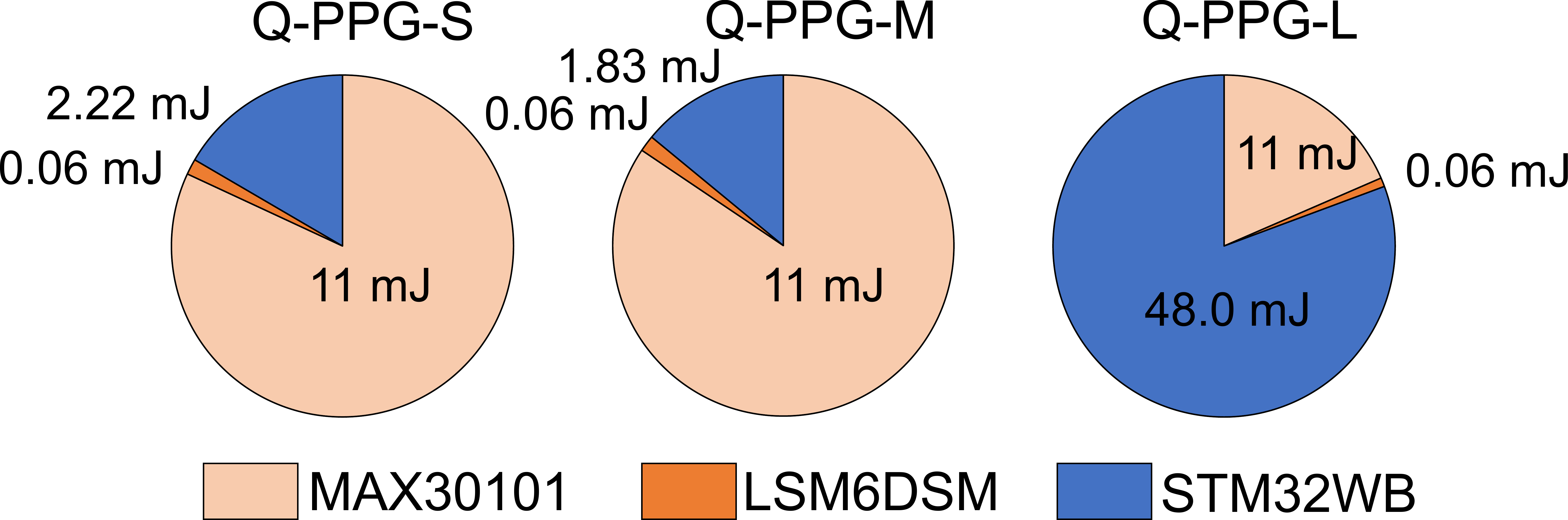}
  \caption{Break-down of the energy consumed in the 2s between two successive HR estimations, including data communication, algorithm execution and waiting time for new data.}
  \label{fig:energy}
  \vspace{-0.5cm}
\end{figure}

In order to compute the power consumption of the entire wearable platform, we then characterized
the two sensors (PPG and accelerometer) and the MCU during data gathering, data communication, and inference, as shown in Table~\ref{tab:power}.
The two sensors i) gather the raw data, ii) store them in internal FIFOs, and iii) send them through I2C/SPI, always consuming a constant on-power of 5.5mW and 0.03mW respectively
%
%
%
The STM32WB stays in \textit{Stop} mode (0.008mW power) between the end of the computation and the moment in which the next window of data is ready to be acquired (gathering phase). 
After that, it goes in \textit{Idle} mode (13.7mW power) during the data communication phase, enabling only the DMA and the SPI/I2C peripherals. Lastly, it goes in \textit{Active} mode (25.0mW power) only to perform inference.
%
%
Note that during the data gathering phase, the power consumption is strongly dominated by the MAX30101 power. 
However, in this work, we do not focus on power-saving techniques for the system's sensing parts, which are set in a default configuration to achieve the sampling rate required by our experiments.

Figure~\ref{fig:energy} reports the energy break-down of the system in a 2s window (the interval between two HR predictions) obtained with the three networks of Table~\ref{tab:deployment}.
SPI/I2C data acquisition and DMA transfers from peripherals to main memory require 15.4ms for both the PPG signal and the acceleration (considering a sampling rate of 32Hz), leading to a stable energy consumption of 11 mJ and 0.06 mJ, respectively.
Conversely, the execution time for inference ranges from 71.6 ms to 1.9 s, always meeting the real-time constraint of 2s.
%
%
In particular, using the Q-PPG-L network as predictor results in an energy consumption of 48.0 mJ, which is 81.3\% of the total consumption of the system.
On the other hand, trading-off a bit of performance for a lighter network, using the Q-PPG-M network, we can reduce the energy consumption for inference to just 1.83 mJ, which is only 14.1\% of the total, with positive effects on battery life, which is critical for embedded and wearable devices.

\section{Conclusions}\label{sec:conclusions}
Executing PPG-based HR tracking on wearable devices is increasingly important in both clinical contexts and daily lives.
However, most research on this task focuses only on maximizing performance, through hand-tuned and often complex algorithms.
To the best of our knowledge, there are no algorithms deployed on wearable-class devices that reach good performance on a large dataset.
In this work, we have tackled this limitation introducing Q-PPG, a new set of quantized deep learning models that span 3 orders of magnitude in memory occupation, with MAEs ranging from a state-of-the-art 4.36 BPM to $\approx$ 20 BPM, on the largest publicly available dataset for this task. All models are derived from a single seed network through the application of a cascade of automatic tools, which progressively either \textit{i)} improve the performance of tracking or \textit{ii)} shrink the model complexity and memory footprint.
We have deployed some of our models on a wearable MCU-based device, demonstrating that they achieve real-time HR tracking with low error, while only contributing to 14.1\% of the total energy consumption of the system, when considering also sensing and communication.

Our development chain is released as open-source at {\texttt{\href{https://github.com/EmbeddedML-EDAGroup/Q-PPG}{https://github.com/EmbeddedML-EDAGroup/Q-PPG}}}.

\bibliographystyle{IEEEtran}

\begin{IEEEbiography}[{\includegraphics[width=1in,height=1.25in,clip,keepaspectratio]{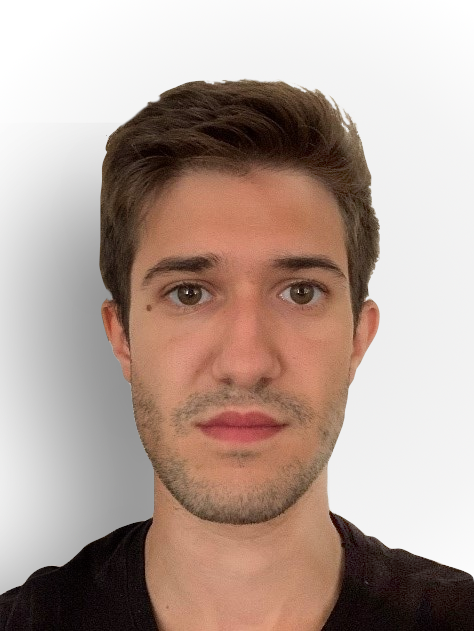}}]{Alessio Burrello}
received his B.Sc and M.Sc degree in Electronic Engineering at the Politecnico of Turin, Italy, in 2016 and 2018.  He is currently working toward his Ph.D. degree at the Department of Electrical, Electronic and Information Technologies Engineering (DEI) of the University of Bologna, Italy.
His research interests include parallel programming models for embedded systems, machine and deep learning, hardware oriented deep learning, and code optimization for multi-core systems.
\end{IEEEbiography}

\begin{IEEEbiography}[{\includegraphics[width=1in,height=1.25in,clip,keepaspectratio]{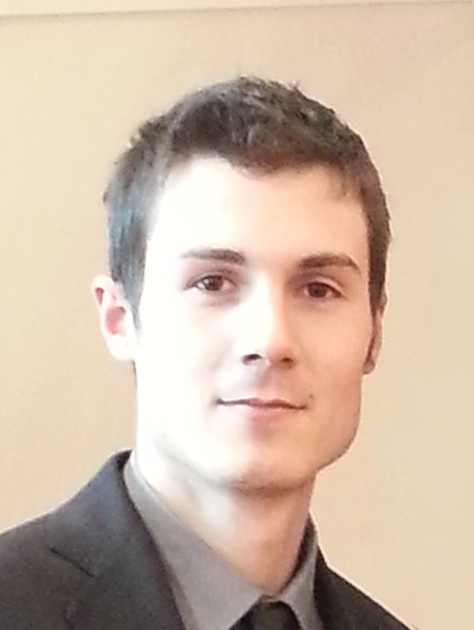}}]{Daniele Jahier Pagliari} received the M.Sc. and Ph.D. degrees in computer engineering from the Politecnico di Torino, Turin, Italy, in 2014 and 2018, respectively. He is currently an Assistant Professor with the Politecnico di Torino. His research interests are in the computer-aided design and optimization of digital circuits and systems, with a particular focus on energy-efficiency aspects and on emerging applications, such as machine learning at the edge.
\end{IEEEbiography}

\begin{IEEEbiography}[{\includegraphics[width=1in,height=1.25in,clip,keepaspectratio]{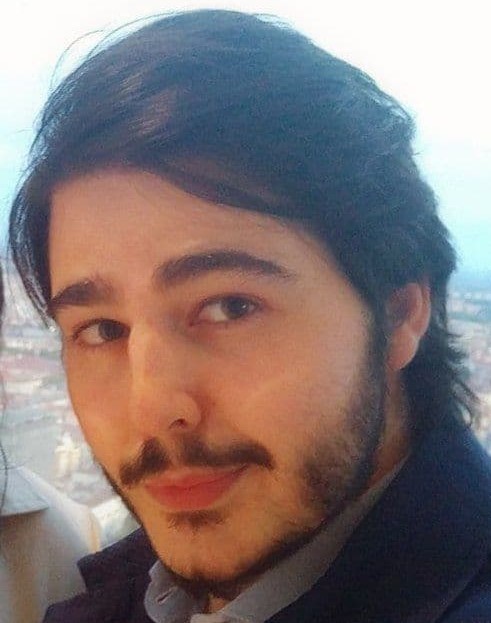}}]{Matteo Risso}
received his B.Sc degree in Physical Engineering and M.Sc degree in Electronic Engineering at the Politecnico di Torino, Italy, in 2018 and 2020. He is currently working toward his Ph.D. degree at Politecnico di Torino, Italy. His research interests include Embedded Machine Learning and Energy-Efficient Embedded Systems.
\end{IEEEbiography}

\begin{IEEEbiography}[{\includegraphics[width=1in,height=1.25in,clip,keepaspectratio]{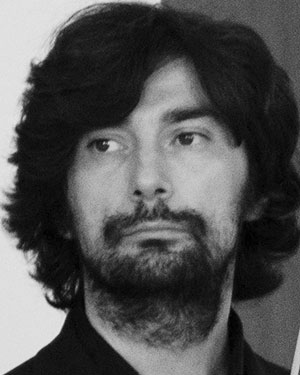}}]{Simone Benatti}
received the Ph.D. degree in electrical engineering and computer science from the University of Bologna, Bologna, Italy, in 2016. He has collaborated with several international research institutes and companies. Previously, he worked for 8 years as an Electronic Designer and R\&D Engineer of electromedical devices. In this ﬁeld, he has authored or coauthored more than 30 papers in international peer-reviewed conferences and journals. His research interests focus on energy efﬁcient embedded wearable systems, signal processing, sensor fusion, and actuation systems. This includes hardware/software codesign to efﬁciently address performance, as well as advanced algorithms.
\end{IEEEbiography}

\begin{IEEEbiography}[{\includegraphics[width=1in,height=1.25in,clip,keepaspectratio]{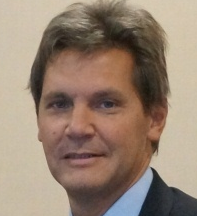}}]{Enrico Macii} is a Full Professor of Computer Engineering with the Politecnico di Torino, Torino, Italy. He holds a Laurea degree in electrical engineering from the Politecnico di Torino, a Laurea degree in computer science from the Università di Torino, Turin, and a PhD degree in computer engineering from the Politecnico di Torino. His research interests are in the design of digital electronic circuits and systems, with a particular emphasis on low-power consumption aspects energy efficiency, sustainable urban mobility, clean and intelligent manufacturing. He is a Fellow of the IEEE.
\end{IEEEbiography}

\begin{IEEEbiography}[{\includegraphics[width=1in,height=1.25in,clip,keepaspectratio]{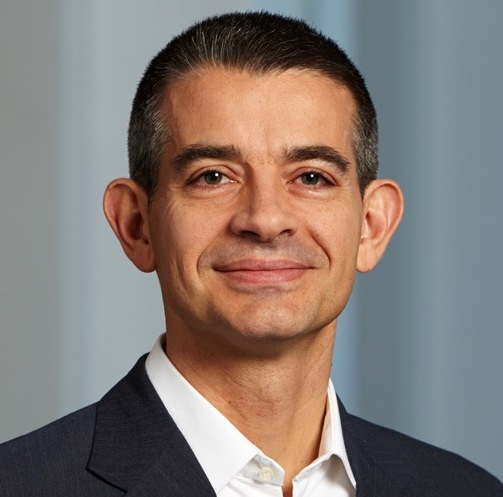}}]{Luca Benini}
is the Chair of Digital Circuits and Systems at ETH Z\"urich and a Full Professor at the University of Bologna.
He has served as Chief Architect for the Platform2012 in STMicroelectronics, Grenoble.
Dr. Benini’s research interests are in energy-efficient systems and multi-core SoC design. 
He is also active in the area of energy-efficient smart sensors and sensor networks. 
He has published more than 1’000 papers in peer-reviewed international journals and conferences, five books and several book chapters. 
He is a Fellow of the ACM and a member of the Academia Europaea.
\end{IEEEbiography}

\begin{IEEEbiography}[{\includegraphics[width=1in,height=1.25in,clip,keepaspectratio]{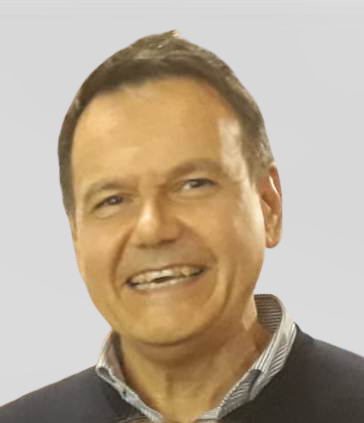}}]{Massimo Poncino} is a Full Professor of Computer Engineering with the Politecnico di Torino, Torino, Italy. His current research interests include various aspects of design automation of digital systems, with emphasis on the modeling and optimization of energy-efficient systems. He received a PhD in computer engineering and a Dr.Eng. in electrical engineering from Politecnico di Torino. He is a Fellow of the IEEE.
\end{IEEEbiography}

\end{document}